\documentclass[11pt,a4paper]{article}
\pdfoutput=1 

\usepackage{jheppub} 
\usepackage[T1]{fontenc}
\usepackage{bbm}
\usepackage{amssymb,amsmath,amsfonts}
\usepackage{mathrsfs,mathtools}
\usepackage[all]{hypcap}
\usepackage{stackrel}
\usepackage{float}
\usepackage{dsfont}
\renewcommand{\thefigure}{\Roman{figure}}

\newcommand{\be}{\begin{equation}}
\newcommand{\ee}{\end{equation}}
\newcommand{\bse}{\begin{subequations}}
\newcommand{\ese}{\end{subequations}}
\newcommand{\ba}{\begin{eqnarray}}
\newcommand{\ea}{\end{eqnarray}}
\renewcommand{\(}{\left(}
\renewcommand{\)}{\right)}

\newcommand{\lk}{\left[}
\newcommand{\rk}{\right]}

\newcommand{\s}{\sigma}

\DeclareMathOperator{\extdm}{d}
\newcommand{\extd}{\extdm \!}
\definecolor{tublue}{rgb}{0,0.2,0.6}
\definecolor{Red}{rgb}{0.7,0,0}
\definecolor{Green}{rgb}{0,0.4,0}
\definecolor{Blue}{rgb}{0,0,0.7}
\definecolor{Violet}{rgb}{0.6,0,0.7}
\definecolor{Cyan}{rgb}{0,0.4,0.5}
\definecolor{Orange}{rgb}{1,0.55,0}
\definecolor{Brown}{rgb}{.37,.24,.06}

\begin{document}

\title{Exploring nonlocal observables in shock wave collisions}

\author[1]{Christian Ecker,}
\author[1]{Daniel Grumiller,}
\author[1]{Philipp Stanzer,}
\author[1]{Stefan A.~Stricker,}
\author[2]{and Wilke van der Schee}

\affiliation[1]{Institut f\"{u}r Theoretische Physik, Technische Universit\"{a}t Wien,\\
Wiedner Hauptstrasse 8-10, A-1040 Vienna, Austria}
\affiliation[2]{Center for Theoretical Physics, Massachusetts Institute of Technology,\\ 
77 Massachusetts Avenue, Cambridge, MA 02139, USA}

\emailAdd{ecker@hep.itp.tuwien.ac.at}
\emailAdd{grumil@hep.itp.tuwien.ac.at}
\emailAdd{pstanzer@hep.itp.tuwien.ac.at}
\emailAdd{stricker@hep.itp.tuwien.ac.at}
\emailAdd{wilke@mit.edu}

\abstract{
We study the time evolution of 2-point functions and entanglement entropy in strongly anisotropic, inhomogeneous and time-dependent ${\cal N}=4$ super Yang--Mills theory in the large $N$ and large 't~Hooft coupling limit using AdS/CFT. On the gravity side this amounts to calculating the length of geodesics and area of extremal surfaces in the dynamical background of two colliding gravitational shockwaves, which we do numerically. We discriminate between three classes of initial conditions corresponding to wide, intermediate and narrow shocks, and show that they exhibit different phenomenology with respect to the nonlocal observables that we determine. 
Our results permit to use (holographic) entanglement entropy as an order parameter to distinguish between the two phases of the cross-over from the transparency to the full-stopping scenario in dynamical Yang--Mills plasma formation, which is frequently used as a toy model for heavy ion collisions.
The time evolution of entanglement entropy allows to discern four regimes: highly efficient initial growth of entanglement, linear growth, (post) collisional drama and late time (polynomial) fall off. Surprisingly, we found that 2-point functions can be sensitive to the geometry inside the black hole apparent horizon, while we did not find such cases for the entanglement entropy.
} 

\preprint{\href{http://www.itp.tuwien.ac.at/Fundamental_Interactions\#2015}{TUW--16--07}, MIT-CTP/4831}


\maketitle

%\noindent
%\cnote{notes by Christian: $\backslash$cnote\{...\}}\\
%\dnote{notes by Daniel: $\backslash$dnote\{...\}}\\
%\pnote{notes by Philipp: $\backslash$pnote\{...\}}\\
%\snote{notes by Stefan: $\backslash$snote\{...\}}\\
%\wnote{notes by Wilke: $\backslash$wnote\{...\}}

\listoffigures

\listoftables

%\newpage
\section{Introduction}\label{se:1}

The gauge/gravity duality has established itself as a valuable tool in the quest for a better understanding of strongly coupled systems. 
In particular it is used  to gain insights into the thermalization process of non-abelian plasmas (such as the quark gluon plasma generated in heavy ion collisions at RHIC and LHC) by studying the gravitational dual of $\mathcal{N}=4$ super Yang--Mills (SYM) theory, a maximally supersymmetric conformal field theory (CFT) in four spacetime dimensions. %, a theory sometimes called the cousin of quantum chromodynamics. 
The equilibration of the field theory is then mapped to black hole formation on the gravity side.
In the last decade there has been considerable progress in setting up  collisions of SYM matter in various scenarios and studying its evolution.

One of the  starting points was the study of perfect fluid dynamics in a boost invariant setting \cite{Janik:2005zt,Janik:2006gp}.
In \cite{Chesler:2008hg} it was possible to study far-from-equilibrium dynamics by numerically solving the full Einstein equations in an anisotropic but otherwise completely homogeneous system. 
Trying to come closer to mimic a  heavy ion collision led to the idea \cite{Janik:2005zt} of colliding delta like gravitational shock waves \cite{Grumiller:2008va,Albacete:2009ji}, which are dual to lumps of energy in the SYM theory moving at the speed of light.
The next step was to make the system anisotropic and inhomogeneous by the collision of  gravitational shock waves which are homogeneous in the transverse direction and have finite width in the longitudinal direction \cite{Chesler:2010bi}. It was found that a hydrodynamic description of the plasma is valid even when the anisotropy is still large \cite{Heller:2011ju}. This onset of  hydrodynamic behavior is now termed hydrodynamization. 
Further advances include radial flow \cite{vanderSchee:2012qj}, the effect of different initial conditions \cite{Casalderrey-Solana:2013aba}, the collision of two black holes \cite{Bantilan:2014sra}, and more \cite{Chesler:2015fpa,Keegan:2015avk,Chesler:2016ceu}.

Now it is even possible to simulate  the collision of two localized lumps  of matter
to mimic off-central nucleus-nucleus \cite{Chesler:2015wra,vanderSchee:2015rta} and proton-nucleus collisions \cite{Chesler:2015bba}.

Despite all the advances one has to keep in mind that in  heavy ion collisions  there are many energy scales involved and to get an accurate understanding  of the thermalization mechanisms involved strong and weak coupling phenomena must be combined. 
One step towards this direction is the combination of different effective descriptions \cite{vanderSchee:2013pia} or by constructing a  semi-holographic framework where the weakly and strongly coupled sector can interact with each other \cite{Iancu:2014ava,Mukhopadhyay:2015smb}.

So far, in most colliding shock wave studies the quantities of interest are local quantities, i.e.\ the components of the energy momentum tensor, such as the energy density or the pressures. This allows to determine if local equilibrium is reached, here understood as the local applicability of hydrodynamics.
In order to gain further insight into the thermalization process the time evolution of  nonlocal quantities, such as various correlation functions (e.g. Wightman function or Feynman propagator),  in coordinate space  need to be considered. This is still a complicated task but two such nonlocal quantities can be obtained relatively easily with the help of the gauge/gravity duality, namely the equal time 2-point function for scalar operators of large conformal weight and entanglement entropy (EE). 
In the context of thermalization these quantities where first computed to study the analog of quenches in conformal field theories \cite{Calabrese:2005in} 
via the collapse of thin shells \cite{AbajoArrastia:2010yt, Balasubramanian:2010ce} in AdS space, where the EE shows universal behavior. 
After the initial short early time epoch the EE grows linearly with time, which is independent of the entangling regions \cite{Liu:2013qca} or the equation of state \cite{Keranen:2014zoa, Keranen:2015fqa}. In these works the EE is a monotonically increasing function that approaches the final equilibrium value from below. 
However, this universal  behavior disappears in more complicated setups. For example, when a radially collapsing scalar field forms a black hole the EE  can be non-monotonic \cite{Buchel:2014gta, Rangamani:2015agy, Abajo-Arrastia:2014fma,daSilva:2014zva,Ecker:2015kna,Bellantuono:2016tkh}.
In anisotropic $\mathcal{N}=4$ SYM the EE and equal time 2-point functions show oscillatory behavior with exponential damping at late times which is given by the lowest quasinormal mode \cite{Ecker:2015kna}.
Analytic progress has been made in \cite{Pedraza:2014moa} where the late-time behavior of two-point functions, Wilson loops and entanglement entropy has been studied perturbatively in a boost-invariant system.

The equal time 2-point function can be obtained from the length of space like geodesics which are anchored to the boundary of anti-de~Sitter (AdS) and extending into the bulk. Although the geodesic approximation is only valid for operators of large conformal weight,
a comparison of the Feynman propagator for a scalar field with conformal dimension $\Delta=3/2$ with the geodesic approximation revealed that qualitatively they show the same behavior \cite{Keranen:2014lna}.
Similarly the holographic entanglement entropy (HEE) can be obtained from the area of extremal surfaces \cite{Ryu:2006bv,Hubeny:2007xt}.

In this work we extend the existing studies by investigating  the time evolution of equal time 2-point functions and HEE in the colliding shock wave geometry for different initial conditions, carefully differentiating between wide, intermediate and narrow shocks, which turn out to have quite different phenomenology. Our results allow to use HEE to distinguish between the phases corresponding to wide or narrow shocks, in a sense that we shall make precise. 

This paper is organized as follows. In Section 2 we introduce the geometry and the different initial conditions. 
The results for the equal time 2-point function and EE are discussed in Sections 3 and 4, respectively. In Section 5 we conclude.

\section{Gravitational shock waves in asymptotically AdS$_5$}\label{se:2}

The holographic setup we consider describes the collision of two sheets of energy having Gaussian shape in their direction of motion and which are homogeneous in the remaining two spatial directions.
These shocks serve as caricatures of two highly Lorentz contracted nuclei  which approach each other at the speed of light and induce non-abelian plasma formation as they collide. 

On the gravity side the corresponding 5-dimensional bulk metric is rotationally invariant and homogeneous in the transverse plane ($x_1,x_2$) but inhomogeneous in the longitudinal direction $y$, which is the direction of motion of the shocks. 
The metric ansatz  in Edding\-ton--Finkelstein coordinates reads
\be\label{metric}
\extd s^2=-A\extd v^2 +S ^2\Big( e^{-2B} \extd y^2 +e^{B} \extd\vec{x}^2 \Big)+2\extd v(\extd r + F\extd y)\;,
\ee
where the functions $A,~S,~B$ and $F$ depend on the holographic coordinate $r$, (advanced) time $v$ and longitudinal coordinate $y$, but are independent from the transversal coordinates $\vec{x}$. 
The equations of motion can be found e.g. in \cite{Chesler:2010bi} and are solved near the boundary by
\bse\label{asymptotic}
\ba
A&=&r^2+2\xi r+ \xi^2-2\partial_v\xi+\frac{a_4}{r^2}+\frac{\partial_v a_4-4\xi a_4}{2r^3}+\mathcal{O}(r^{-4})\\
B&=&\frac{b_4}{r^4}+\frac{15\partial_v b_4+2\partial_y f_4 -60\xi b_4}{15r^5}+\mathcal{O}(r^{-6})\\
S&=&r+\xi-\frac{4\partial_y f_4 + 3\partial_v a_4}{60r^4}+\mathcal{O}(r^{-5})\\ 
F&=&\partial_y\xi +\frac{f_4}{r^2}+\frac{4\partial_v f_4+\partial_y a_4 -10\xi f_4}{5r^3}+\mathcal{O}(r^{-4}),
\ea
\ese
where $\xi(v,y)$ encodes  the residual diffeomorphism freedom $r\to r+\xi(v,y)$. 
It is possible, though not necessarily numerically convenient, to choose $\xi=0$. 

As usual the normalizable modes $a_4(v,y)$, $b_4(v,y)$ and $f_4(v,y)$ are undetermined by the near-boundary expansion and require a solution of the full bulk dynamics. These coefficients in the asymptotic expansion determine the expectation value of the conserved and traceless stress energy tensor in the dual field theory \cite{deHaro:2000xn}
\be\label{EMT}
\langle T^{\mu\nu}\rangle =\frac{N_c^2}{2\pi^2}
\begin{pmatrix}
  \mathcal{E}                & \mathcal{\mathcal{S}}                & 0                                & 0                               \\
  \mathcal{\mathcal{S}}      & \mathcal{\mathcal{P}}_\parallel      & 0                                & 0                               \\
  0                          & 0                                    & \mathcal{\mathcal{P}}_\perp      & 0                               \\
  0                          & 0                                    & 0                                & \mathcal{\mathcal{P}}_\perp     \\
 \end{pmatrix}
\ee
where 
\be
\mathcal{E}=-\frac{3}{4}a_4\qquad \mathcal{P}_\parallel = -\frac{1}{4}a_4-2b_4\qquad \mathcal{P}_\perp = -\frac{1}{4}a_4 +b_4\qquad \mathcal{S} = -f_4\,.
\ee

%%%%%%%%%%%%%%%%%%%%%%%%%%%%%%%%%%%%%%%%
\subsection{Initial conditions}\label{se:2.1}
%%%%%%%%%%%%%%%%%%%%%%%%%%%%%%%%%%%%%%%%

The pre-collision  geometry  describing  two shocks moving in $\pm \tilde{y}$-direction can be written down in Fefferman-Graham coordinates ($\tilde r$, $\tilde t$, $\tilde y$, $\vec{\tilde x}$)  as follows \cite{Janik:2005zt}
\be
\label{ansatzFG}
\extd s^2=\tilde{r}^2\eta_{\mu\nu}\extd \tilde{x}^\mu \extd \tilde{x}^\nu +\frac{1}{\tilde r^2}\Big(\extd \tilde{r}^2+h(\tilde{t}+\tilde{y})(\extd \tilde{t}+\extd \tilde{y})^2+h(\tilde{t}-\tilde{y})(\extd \tilde{t}-\extd \tilde{y})^2\Big),
\ee
where $\eta_{\mu\nu}$ denotes the usual 4-dimensional Minkowski boundary metric %located at $\tilde r=\infty$ 
and $h(\tilde{t}\pm\tilde{y})$ is  an arbitrary function for which we choose a Gaussian of width $\omega$ and amplitude $\mu^3$
\be
\label{IC}
h(\tilde{t}\pm\tilde{y})=\frac{\mu^3}{\sqrt{2\pi\omega^2}}e^{-\frac{(\tilde{t}\pm\tilde{y})^2}{2\omega^2}}\,.
\ee
In this gauge the non-vanishing components of the energy momentum tensor read
\be
\tilde T^{\tilde t \tilde t}=\tilde T^{\tilde y \tilde y}=h(\tilde{t}-\tilde{y}) + h(\tilde{t}+\tilde{y})\qquad \tilde T^{\tilde t \tilde y}=h(\tilde{t}-\tilde{y}) - h(\tilde{t}+\tilde{y})
\ee
and describe two lumps of energy with maximum overlap at $\tilde t=0$.
At early times $\tilde t \ll -w$, when the shocks $h(\tilde{t}\pm\tilde{y})$ have negligible overlap, the line-element (\ref{ansatzFG}) is close to an exact solution to  Einstein's equations, but around $\tilde t=0$ their dynamics can only be computed numerically.

We do this for three  different initial conditions $h_{n,i,w}(\tilde y)$ describing qualitatively different situations that we shall refer to as narrow, intermediate and  wide shocks,  where in all cases the initial position of the shocks is at $\tilde{y}_0=\pm7/4$. 
For the width of the shocks we take $\omega_{n,i,w}=0.1,~0.25,~0.5$ and we will display all our results in units of $\mu$.

For the numerical  evolution scheme the initial data needs to be transformed to Edding\-ton--Fin\-kel\-stein coordinates $(r,v,y,x_1,x_2)$ by solving for radially infalling  null geodesics in the background  (\ref{ansatzFG}), leading to ordinary differential equations, which  are solved   for appropriate boundary conditions at the boundaries of the radial domain.
We omit a discussion of the numerical details concerning this coordinate transformation and the subsequent evolution and refer the reader to \cite{vanderSchee:2014qwa,Chesler:2013lia}, where the full procedure is explained.

%%%%%%%%%%%%%%%%%%%%%%%%%%%%%%%%%%%%%%%%%%%%%%%
\subsection{Evolution of the energy momentum tensor}\label{se:2.2}
%%%%%%%%%%%%%%%%%%%%%%%%%%%%%%%%%%%%%%%%%%%%%%%%

\begin{figure}
%\begin{center}
\hspace{-1.cm}
\includegraphics[scale=.22]{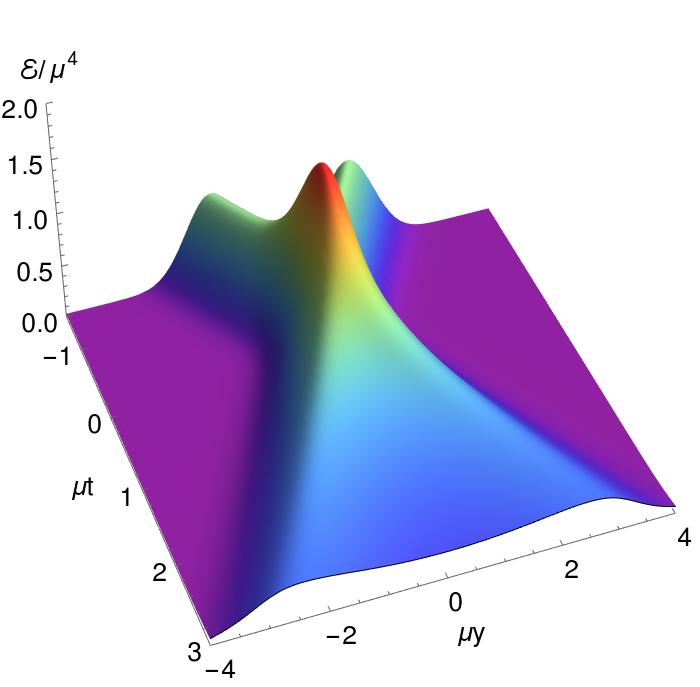}$\;$\includegraphics[scale=.22]{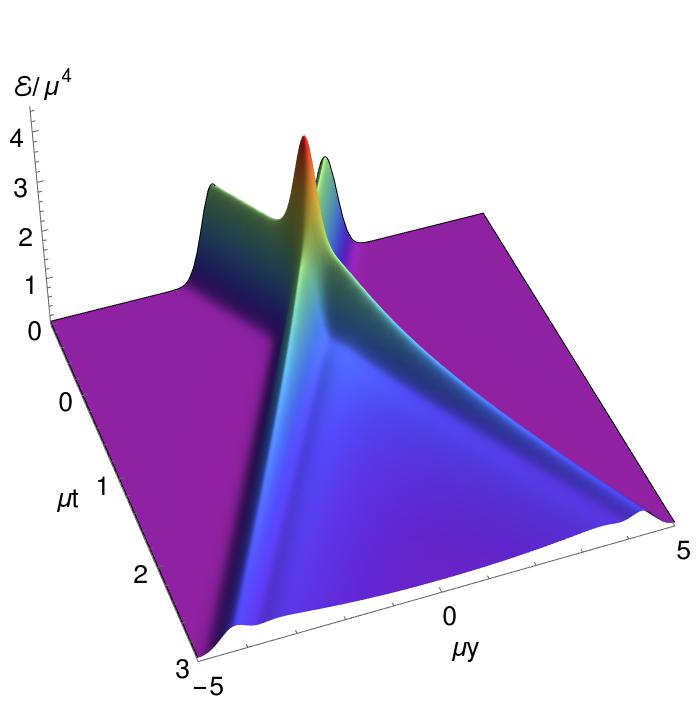}$\;$\includegraphics[scale=.22]{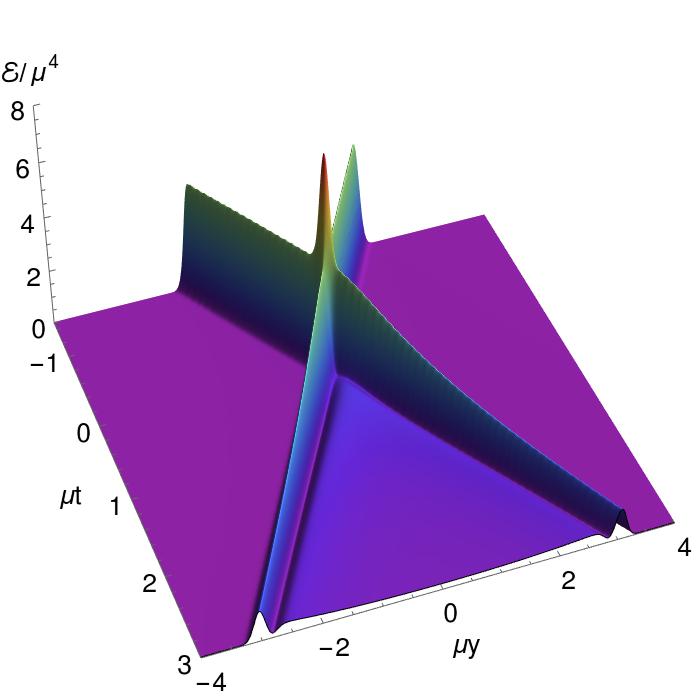}
\caption[Anisotropy function and pressures.]{\label{EMTevolution} 
Evolution of the energy density $\epsilon/\mu^4$ as a function of time $t$ and longitudinal coordinate $y$ for wide, intermediate and narrow shocks (from left to right).}
%\end{center}
\end{figure}

The time evolution of the energy momentum tensor for colliding shocks has been studied extensively in \cite{Chesler:2010bi,Casalderrey-Solana:2013aba,Casalderrey-Solana:2013sxa,Chesler:2013lia}.
In Fig.~\ref{EMTevolution} we show the evolution of the energy density $\mathcal{E}(t,y)$ extracted from the numerical evolution for the different initial conditions stated above.
As discussed in \cite{Casalderrey-Solana:2013aba} the energy density behaves qualitatively different in collisions of narrow shocks and in those of wide shocks. This cross-over is not only of academic interest, but also for applications, since it was argued that the narrow shocks describe more adequately the situation at LHC, while the wide shocks are more suitable for RHIC  \cite{Casalderrey-Solana:2013aba} (see also \cite{vanderSchee:2015rta}). We list below some relevant properties that differ between wide and narrow shocks:
\begin{itemize}
\item Narrow shocks exhibit transparency, in the sense that they pass through each other and, even though their shape gets altered and they decay, they continue to move at the speed of light after the collision. By contrast, wide shocks realize a full-stopping scenario, in the sense that the energy density is localized mostly in the central region after the collision, and the shocks themselves not only change their shape but also get slowed down. Wide shocks then lead to initial conditions for hydrodynamics where all velocities are close to zero, i.e. there is a hydrodynamical explosion in close similarity to the Landau model of heavy ion collisions \cite{Landau:1953gs}.
\item Narrow shocks can yield regions of negative energy density after the collision right behind the original shocks on the lightcone. Curiously, this region does not admit a local restframe \cite{Arnold:2014jva}, but also does not violate general principles of quantum field theory, such as the averaged null energy condition \cite{Ford:1999qv}. At $y=0$ after the shocks pass through each other, the energy density grows at early times as $\mathcal{E}=2 \mu^6 t^2+\mathcal{O}(t^5)$, which implies pressures equal to $\mathcal{\mathcal{P}}_\parallel/\mathcal{E}=-3$ and $\mathcal{\mathcal{P}}_\perp/\mathcal{E}=2$.
This feature was first observed for $\delta$-like shockwaves analytically \cite{Grumiller:2008va} and then numerically for sufficiently narrow Gaussian profiles \cite{Casalderrey-Solana:2013aba}. 
By contrast, for the wide shocks the energy density and pressures remain positive everywhere.

\end{itemize}
Given the substantial differences in local observables one may expect that the characteristic features for narrow and wide shocks also show up in nonlocal observables, like 2-point functions and HEE. In the remainder of this work we verify this expectation by explicit computations, starting with the 2-point functions in the next section.

%%%%%%%%%%%%%%%%%%%%%%%%%%%%%%%%%%%%%%%%%%%
%%%%%%%%%%%%%%%%%%%%%%%%%%%%%%%%%%%%%%

\section{Two-point functions}\label{se:3}

Within AdS/CFT the equal time 2-point function of operators $\mathcal{O}$ with large conformal weight $\Delta$ can be computed from the length $L$ of spacelike geodesics in the bulk geometry \cite{Balasubramanian:1999zv, Festuccia:2005pi} via
\be
\langle \mathcal{O}(t, \vec{x})\mathcal{O}(t, \vec{x}')\rangle=\int  \mathcal{D P}\, e^{i \Delta \mathcal{L(\mathcal{P})} }\approx \!\!\!\sum_{\textrm{\tiny geodesics}}\!\!\! e^{-\Delta L_g}\approx e^{-\Delta L}\;.
\ee
In asymptotically AdS the length of a geodesic which is attached to the boundary is infinite and a regularization scheme must be adopted.
A natural way to regularize is to subtract the length $L_0$ of a geodesic in AdS corresponding to the vacuum value of the correlator 
\be
L_{\textrm{\tiny reg}}=L-L_0 \;.
\ee

For illustrative purposes we set $\Delta=1$ when we display our results which is the same as interpreting $L_{\textrm{\tiny reg}}$ to be given in units of $\Delta$.
Thus, the two point functions we compute are defined as follows
\be\label{2pfreg}
\langle \mathcal{O}(t, \vec{x})\mathcal{O}(t, \vec{x}')\rangle_{\textrm{\tiny reg}}= e^{-L_{\textrm{\tiny reg}}}\;.
\ee

% This is a natural choice because at late times the 2-point function is time independent. 
In order to obtain the geodesic length we solve the geodesic equation numerically with a relaxation algorithm which iteratively relaxes an initial guess to the true solution.
For a detailed description of the relaxation algorithm we refer the interested reader to \cite{Ecker:2015kna}.
%%%%%%%%%%%%%%%%%%%%%%%%%%%%%%%%%%%%%%%%%%%%
\subsection{Geodesics in the shock wave geometry}\label{se:3.1}
%%%%%%%%%%%%%%%%%%%%%%%%%%%%%%%%%%%%%%%%%%%%

For simplicity we restrict our attention to geodesics that only  extend  along the $y$-direction  and not along the transverse directions ($x_1,x_2$), i.e.\ we consider geodesics in the three dimensional bulk-subspace 
\ba\label{submetric}
\extd s^2_y&=&  -A \extd v^2-\frac{2}{z^2}  \extd z \extd v +2 F\extd y\extd v +S^2 e^{-2B} \extd y^2,
\ea
where   $z=1/r$.
To find these geodesics we solve the (non-affine) geodesic equation
\be
\ddot X^\mu + \Gamma^\mu{}_{\alpha\beta} \dot X^\alpha \dot X^\beta = -J \dot X^\mu,
\label{eq:geo}
\ee
%subject to the following boundary conditions at some IR-cutoff $z=z_{\textrm{\tiny cut}}$
subject to the following boundary conditions at $z=0$
\be
X^\mu(\pm1)\equiv(V(\pm1),Z(\pm1),Y(\pm1))=(t,0,\pm l/2),
\label{eq:para}
\ee
where $X^\mu(\sigma)$ are the embedding functions of the geodesic and dots denote derivatives with respect to the non-affine parameter $\sigma\in [-1,1]$.
 The quantity  $J=\frac{\extd^2\tau}{\extd\sigma^2}/\frac{\extd\tau}{\extd\sigma}$ denotes the Jacobian of the reparametrization from the affine parameter $\tau$, defined by $(\frac{\mathrm{d}X}{\mathrm{d}\tau})^2=1$, to $\sigma$. The boundary time and separation for which the geodesics are computed are denoted by $t$ and $l$ respectively. The fictitious viscous force provided by the Jacobian $J$ helps with the numerics, resulting in better convergence of the relaxation algorithm. 

Working in  asymptotically AdS makes it natural to choose as an initial guess
a geodesic in pure AdS 
\be\label{adsmetric}
\extd s_0^2=\frac{1}{z^2}\(-\extd v^2-2 \extd z \extd v + \extd y^2\)\;,
\ee
which can be written as 
\ba\label{ansatz2PF}
Z_0(\s)=\frac{l}{2}\big(1-\s^2\big)\;,\qquad Y_0(\s)=\frac{l}{2}\big(\s \sqrt{2-\s^2}\big)\;,\qquad V_0(\s)=t-Z_0(\s).
\ea
In this parametrization the affine parameter is given by $\tau(\sigma)\!=\!\mp\mathrm{arctanh}\Big(\sigma \sqrt{2-\sigma^2}\Big)$ from which the Jacobian needed in (\ref{eq:geo}) can be computed 
\be\label{Jacobian2PF}
J(\sigma)=\frac{\extd^2\tau}{\extd\sigma^2}\Big/\frac{\extd\tau}{\extd\sigma}=\frac{5\sigma-3\sigma^3}{2-3\sigma^2+\sigma^4}\,.
\ee
We assume the boundary separation to be centered around $y=0$. Describing off-central geodesics requires some straightforward modifications of our formulas. 

The bulk part of the geodesic length, which is the contribution from $z>z_{\textrm{\tiny cut}}$, follows from integrating the line elements (\ref{submetric}) and (\ref{adsmetric}) 
\bse\label{Lbulk}
\ba
L^{\textrm{\tiny bulk}}&=&\int_{\s_-}^{\s_+} \extd\s \sqrt{-A \dot{V}^2-\frac{2}{Z^2}\dot{Z}\dot{V}+2F\dot{V}\dot{Y} +S^2 e^{-2B}\dot{Y}^2},\\
L^{\textrm{\tiny bulk}}_0&=&\int_{\s_-}^{\s_+} \extd\s \frac{1}{Z_0}\sqrt{- \dot{V_0}^2-2 \dot{Z}_0 \dot{V}_0 + \dot{Y}_0^2},
\ea
\ese
where the metric functions $(A,B,S,F)$ have  to be evaluated along the geodesic $X^\mu(\s)$. 
In order to realize an IR-cutoff at a given value $z_{\textrm{\tiny cut}}$ the range of the non-affine parameter $\sigma\in [\sigma_-,\sigma_+]$ has to be bounded by
\be
\s_\pm=\pm\sqrt{1-\frac{2 z_{\textrm{\tiny cut}}}{l}}.
\ee
The near boundary part of the geodesic length, which is the contribution from $0\le z \le z_{\textrm{\tiny cut}}$, can be extracted form the near boundary solution of the geodesic equation.
Near $z=0$ the embedding functions and the Jacobian can be expressed in terms of a power series in $z$
\be\label{asympAnsatz}
 Z(z)=z,\qquad V(z)=\sum_{n=1}^{n_{\textrm{\tiny max}}}v_n z^n,\qquad Y(z)=\sum_{n=1}^{n_{\textrm{\tiny max}}}y_n z^n,\qquad J(z)=\sum_{n=1}^{n_{\textrm{\tiny max}}}j_n z^{n-2}\,,
\ee
In Appendix \ref{app:1} we give the explicit expressions for the expansion of the metric that we have used.
The coefficients $(t_n,y_n,j_n)$ in Eq.~(\ref{asympAnsatz}) can be computed by solving the geodesic equation order by order in $z$, which leads to the following expressions
\bse\label{asympSolution2PF}
\ba
 Z(z) & = & z \\
 V(z) & = & v_0-z+v_2 z^2+\left(v_2 y_2^2-v_2^3\right) z^4+O\left(z^5\right) \\
 Y(z) & = & \frac{l}{2}+y_2 z^2+\left(y_2^3-v_2^2 y_2\right) z^4+O\left(z^5\right) \\
 J(z) & = & \frac{1}{z}+\left(4 v_2^2-4 y_2^2\right) z+O\left(z^5\right) \\
\ea
\ese
Here we fixed the leading coefficients by the boundary conditions (\ref{eq:para}), but the coefficients  $v_2$ and $y_2$ cannot be determined by
a near boundary expansion. This is analogous to the normalizable modes of the metric, which are also sensitive to the full bulk geometry.
The pure AdS solution is given by
\bse\label{asympGeo0}
\ba
 Z_0(z)&=&z,\\
 V_0(z)&=&t_0-z,\\
 Y_0(z)&=&\pm \sqrt{(l/2)^2-z^2}\nonumber\\
       &=&\pm \Big( \frac{l}{2}-\frac{z}{l}-\frac{z^4}{l^3}\Big)+\mathcal{O}(z^{6}),\\
 J_0(z)&=&\frac{1}{z}-\frac{4}{l^2}z-\frac{16}{l^4}z^3+\mathcal{O}(z^{5})\;.
\ea
\ese
which hence has $v_2=0$ and $y_2=\mp 1/l$. We can now compute the near boundary expansion of the geodesic length, which for one branch is given by
\bse\label{bdryL}
\ba
L^{\textrm{\tiny bdry}}-L_0^{\textrm{\tiny bdry}}&=& \int_{0}^{z_{\textrm{\tiny cut}}} \extd z  \left(-\frac{2}{l^2}-2 v_2^2+2 y_2^2\right)z\nonumber\\
&+&\left(-\frac{a_4}{2}-\frac{6}{l^4}-12 v_2^2 y_2^2+6
   v_2^4+6 y_2^4\right)z^3 +O\left(z^5\right),
\ea
\ese
where the leading AdS divergent $\tfrac{1}{z}$ term nicely cancels out.
The regularized geodesic length $L_{\textrm{\tiny reg}}$, which we need to evaluate Eq.~(\ref{2pfreg}), is the sum of the bulk contribution and the near boundary contribution 
\footnote{In practise we do not compute the near boundary term, as the extraction of $v_2$ and $y_2$ would be numerically as hard as taking a small enough $z_{cut}$
such that this term is small. We have included this formula for completeness, and will later see that a similar procedure does work for entanglement entropy.}  
\be\label{Lreg}
L_{\textrm{\tiny reg}}=(L^{\textrm{\tiny bulk}}-L_0^{\textrm{\tiny bulk}})+(L^{\textrm{\tiny bdry}}-L_0^{\textrm{\tiny bdry}})\;.
\ee
When using Eq.~(\ref{Lreg}) to evaluate Eq.~(\ref{2pfreg}) numerically one has to ensure that the results are, to some required accuracy, independent of the discretization and the cutoff.
 We require this accuracy to be of the same order as the maximal residual ($=10^{-5}$) we allow in the geodesic equation and below which we stop to iterate the relaxation procedure.
We checked the convergence of the 2-point function with the gridsize in the range from 50 up to 400 gridpoints and find that for more than $200$ gridpoints the change is smaller than $\mathcal{O}(10^{-5})$ which is the same order as the allowed residual (see Appendix \ref{app:2}). 
Sample checks are presented in Appendix \ref{app:2}, where only a mild cutoff dependence of $\mathcal{O}(10^{-5})$ is obtained for a range $z_{\textrm{\tiny cut}}=[0.01,0.1]$, which is again of the same order as the allowed residual. 
Based on this analysis we choose $200$ gridpoints to discretize our geodesics and set $z_{\textrm{\tiny cut}}=0.075$ in all our calculations.

%%%%%%%%%%%%%%%%%%%%%%%%%%%%%%%%%%%%%%%%%%%%
\subsection{Evolution of two-point functions}\label{se:3.2}
%%%%%%%%%%%%%%%%%%%%%%%%%%%%%%%%%%%%%%%%%%%%
In this section we present our numerical results for 2-point functions in holographic shock wave collisions.
Before we discuss the actual results let us start with some remarks regarding the computational domain used in these simulations.
As input for the relaxation algorithm we provide numerical results of the shock wave metric in a finite domain in $(t,y,z)$.  This computational domain, in which we can solve the geodesic problem, is bounded by $\mu t\in[-1.5,6]$, $\mu y\in[-5,5]$, where in the radial coordinate we have chosen the apparent horizon as a natural bound $z\in[0,1.08 z_{\textrm{\tiny AH}}]$.
That means whenever we display geodesics which reach beyond this radial domain, which can happen as we discuss below, an extrapolated version of the metric is used\footnote{
For the narrow shocks the computational domain does not reach behind the horizon, so there extrapolation is always used (note that the fact that the geodesic crosses the horizon or not is not affected by this extrapolation).}.
For a given choice of boundary conditions $(\mu t,\mu l)$ the final shape of the geodesic in the bulk is a priori unknown, i.e.\ initially we do not know if the geodesic resides entirely within or extends beyond the computational domain in which the metric is known. Therefore  finding a feasible set of parameters $(\mu t,\mu l)$ for a given computational domain requires some trial and error. 
The geodesics bend back in advanced time as they reach into the bulk, leaving the computational domain for too early boundary times. Therefore we can display our results only in a finite time near the collision time $t=0$ where all geodesics with different boundary separation lie in the computational domain.
All these points apply accordingly to the EE simulation.

\begin{figure}
%\begin{minipage}[t][15cm]{1.0\textwidth}
\begin{minipage}[t]{0.4\textwidth}
\vspace{-0.2cm}
\hspace{-0.2cm}
\includegraphics[width=\textwidth]{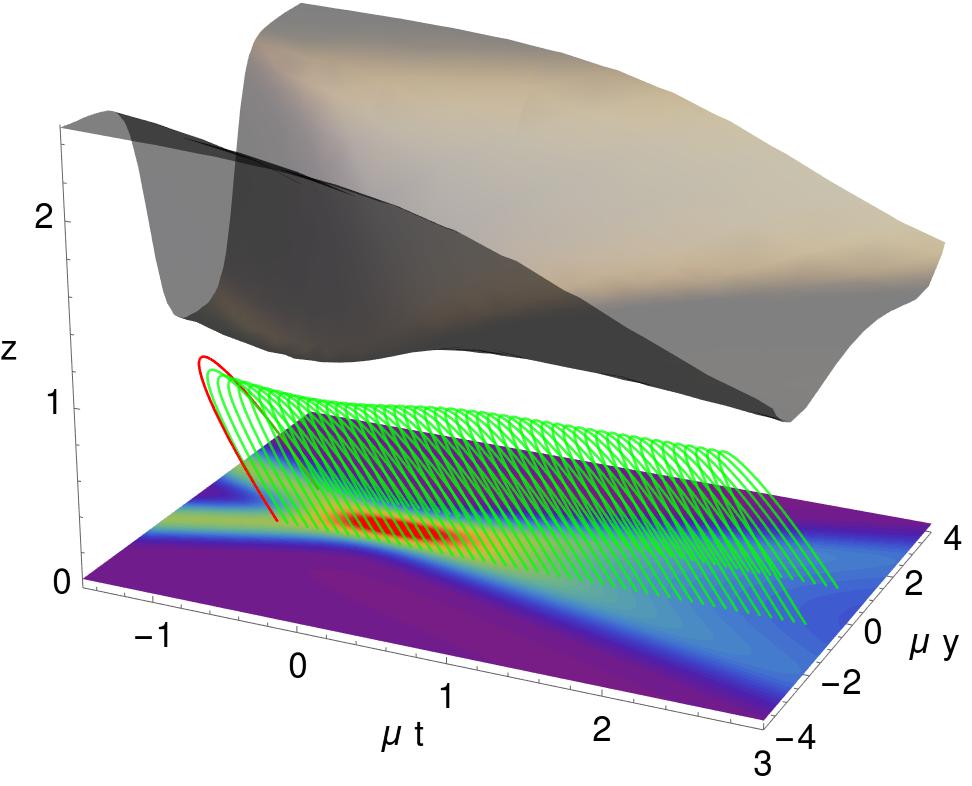}
\end{minipage}
\begin{minipage}[t]{0.4\textwidth}
\vspace{0.8cm}\includegraphics[width=\textwidth]{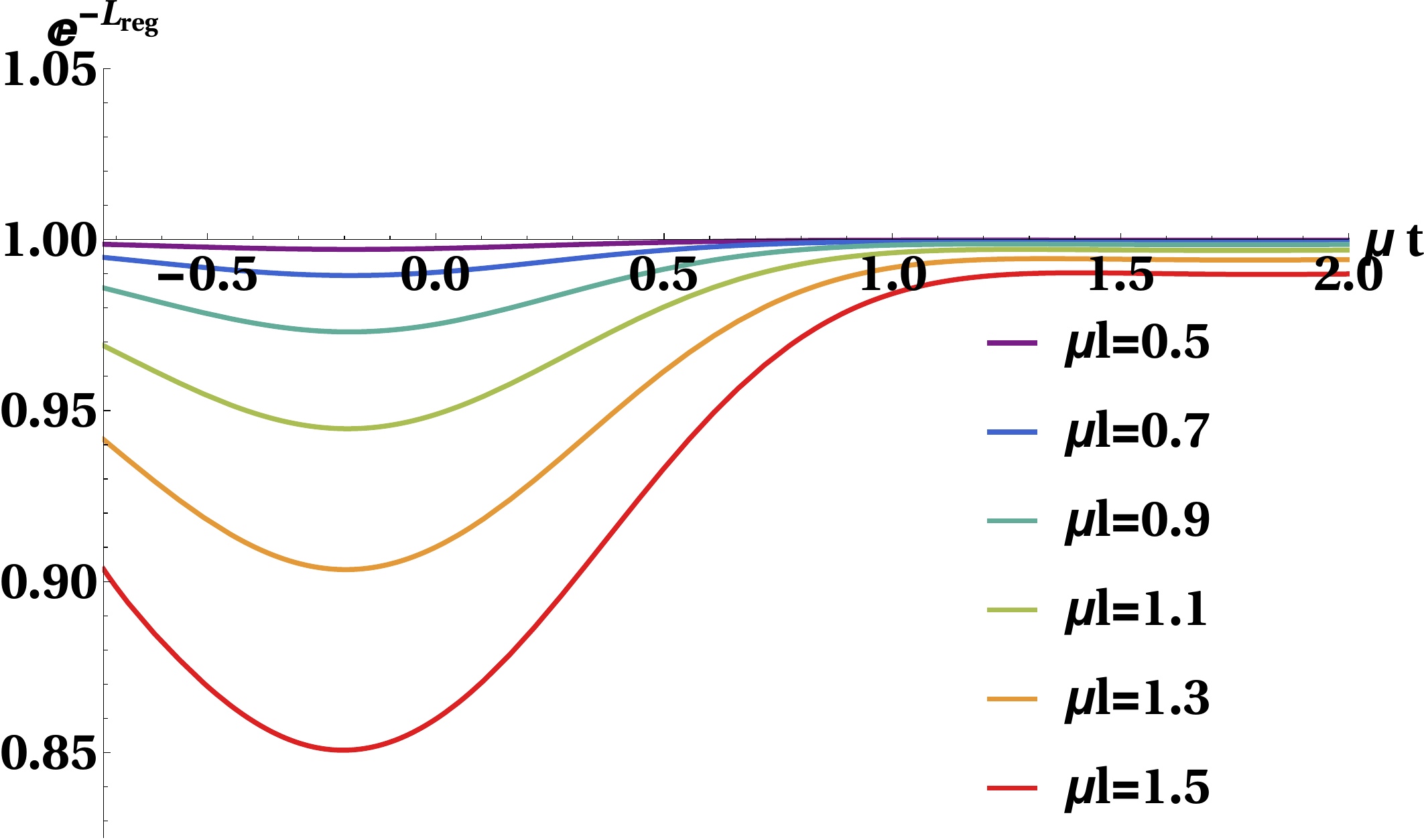}
\end{minipage}
\begin{minipage}[t]{0.4\textwidth}
\vspace{-0.0cm}
\hspace{-0.2cm}
\includegraphics[width=\textwidth]{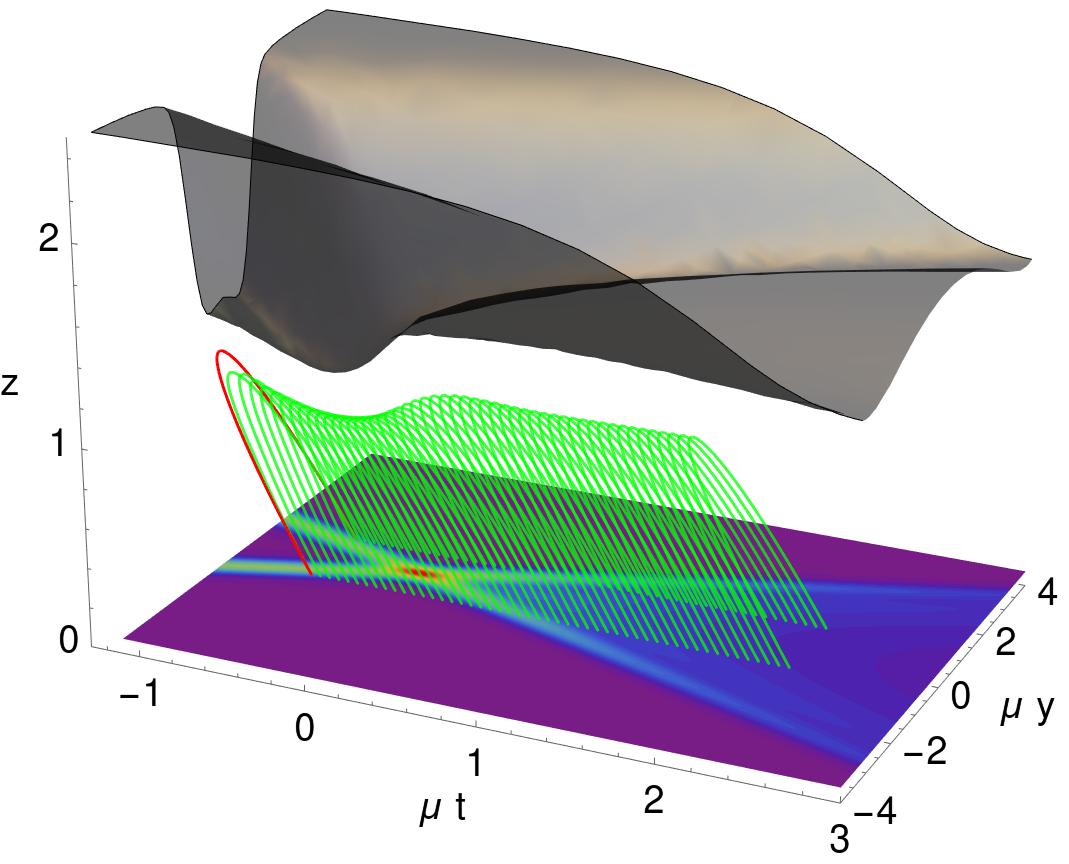}
\end{minipage}
\begin{minipage}[t]{0.4\textwidth}
\vspace{0.8cm}\includegraphics[width=\textwidth]{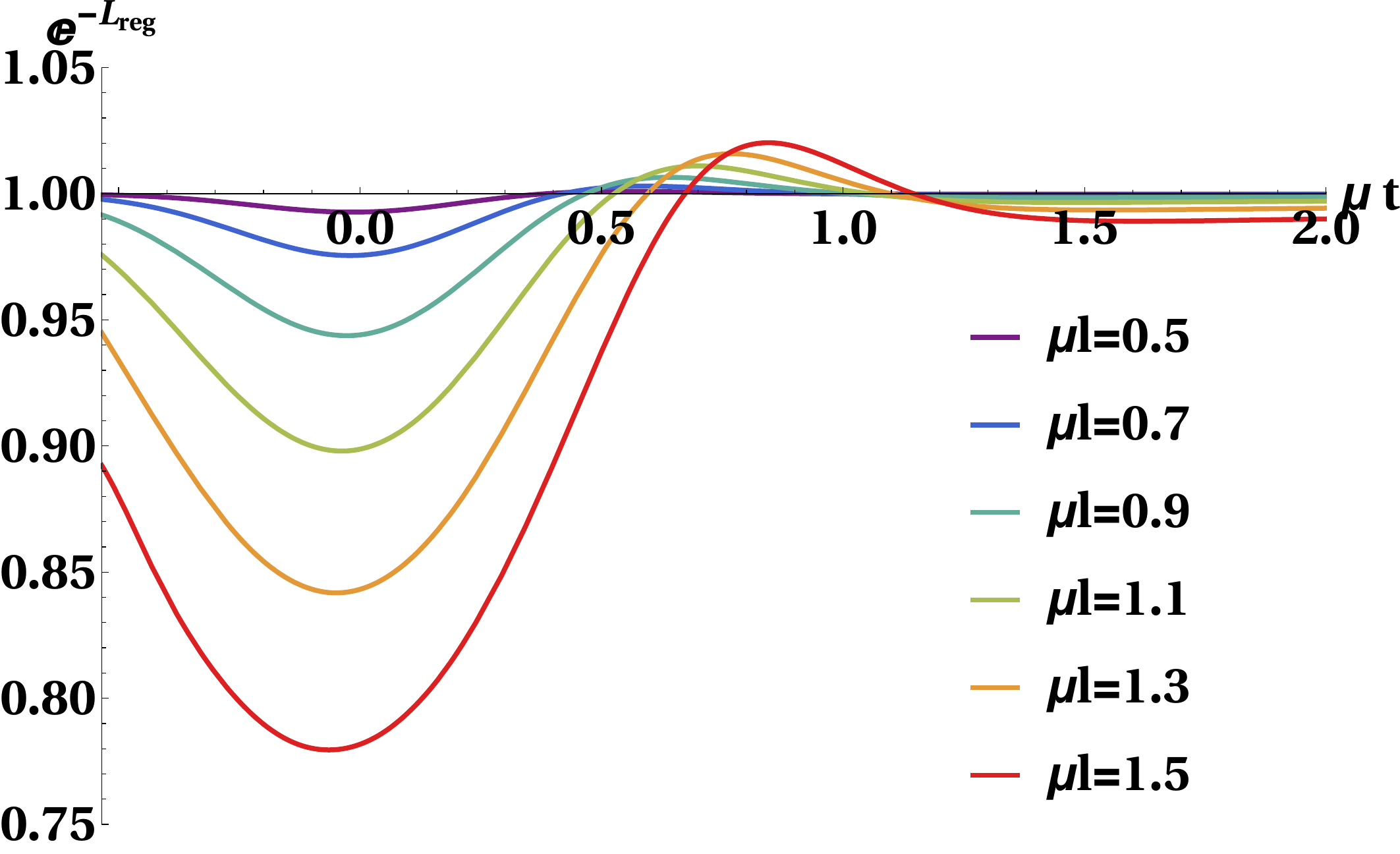}
\end{minipage}
\begin{minipage}[t]{0.4\textwidth}
\vspace{-0.0cm}
\hspace{-0.2cm}
\includegraphics[width=\textwidth]{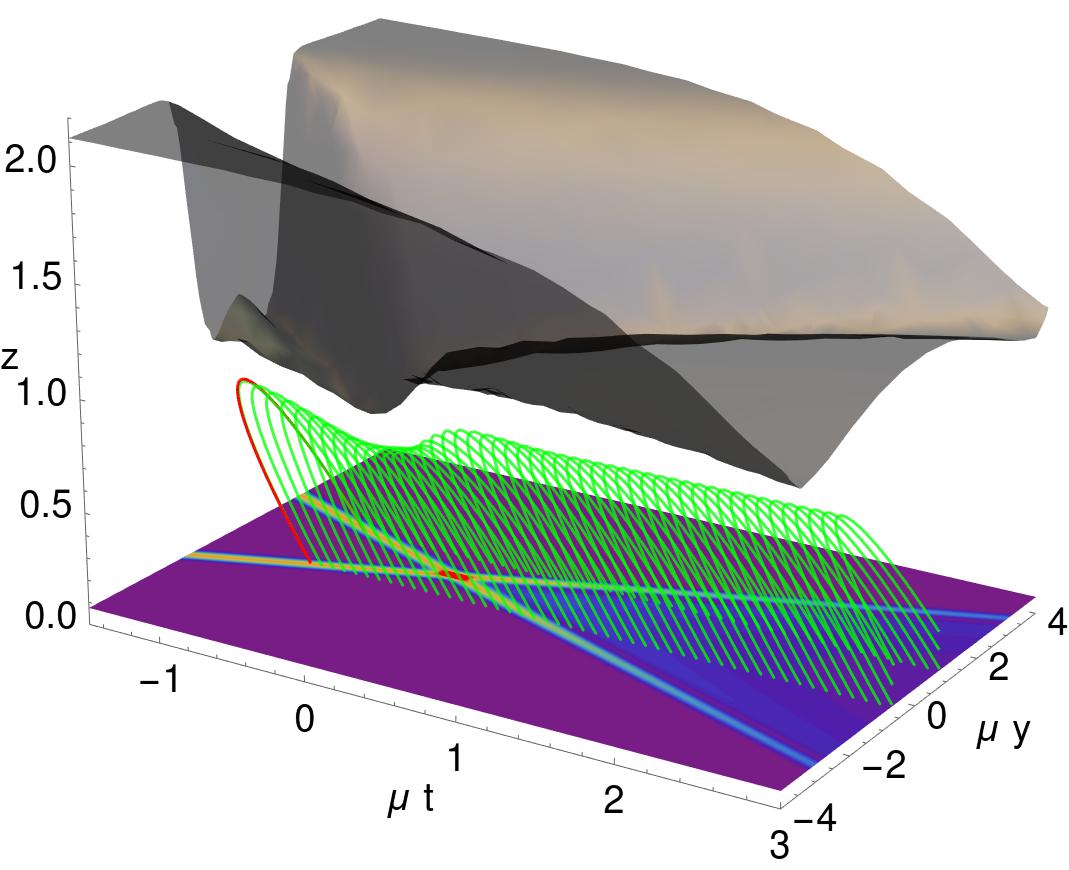}
\end{minipage}
\begin{minipage}[t]{0.4\textwidth}
\vspace{0.8cm}
\hspace{2.8cm}
\includegraphics[width=\textwidth]{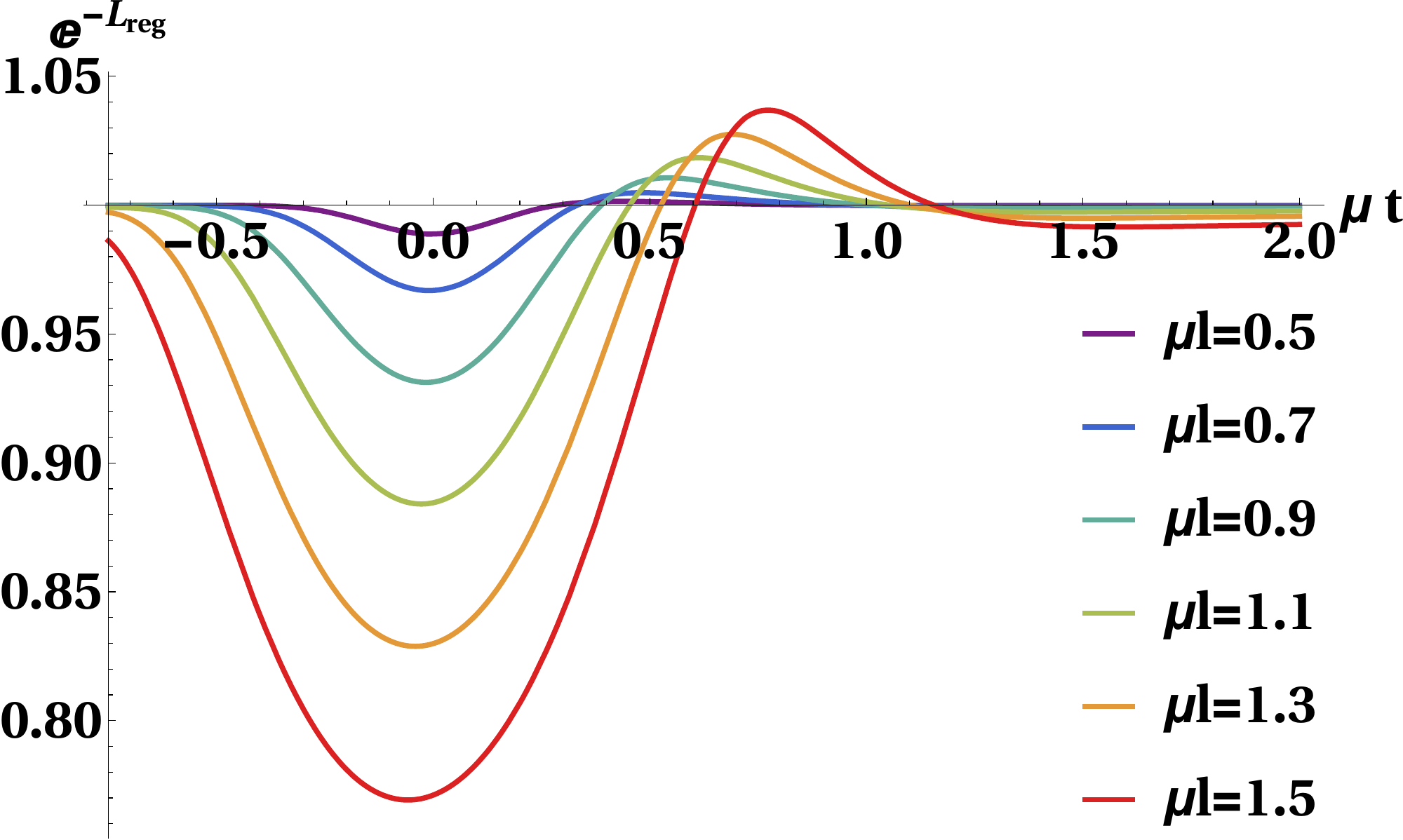}
\end{minipage}
\caption[Geometrical setup and 2-point function (time-evolution).]{Left: Summary of the geometrical setup.
The black surfaces represent the radial position $z_{\textrm{\tiny AH}}(t,y)$ of the apparent horizon; red curves are AdS geodesics used for the initialization, the  green lines are geodesics ($\mu l=1.5$) for various time steps and at $z=0$ we show a density plot of the energy density for wide, intermediate and narrow shocks (top to bottom).
Right:  Corresponding evolution of the 2-point function for different boundary separation $\mu l$.}\label{2PFfixed}
%\end{minipage}
\end{figure}

For the time evolution it is of advantage, after using the pure AdS geodesic at the initial time, to use the previous solution to initialize the next time step. This approach turns out to be numerically extremely efficient and the relaxation algorithm reveals its full power, since in most cases the result at a given time is an excellent estimate at the next time step. A time step of $\Delta t=0.1$ allows to resolve nicely the shape of the 2-point function and reduces the required number of iterations almost to a minimum. Usually two iterations are sufficient to achieve relative residuals in the geodesic equation which are $<10^{-5}$ and in many cases even one or two orders smaller. 

We follow the same logic when we compute the evolution in the boundary separation, where this approach is not only numerically efficient but also crucial to reach large separations. Undeformed ansatz geodesics of large separation typically reach far beyond the radial domain and finding the true solution using such geodesics to initialize the relaxation inevitably fails. We circumvent this problem by initializing with an ansatz of small separation ($\mu l=0.2$), which comfortably resides within the computational domain. Then we increase step by step the boundary separation and use the result for a given separation as ansatz for the next separation step. By using a step size of $\Delta l=0.1$ we can nicely resolve the shape of the 2-point function and the relaxation usually converges again after two iterations. Since the relaxed geodesics are typically strongly deformed in direction away from the apparent horizon, i.e.\ the upper bound of the radial domain, we can reach separations which were inaccessible by simply relaxing the corresponding ansatz geodesic.

We like to discuss first the results from the time evolution before we go to the evolution in the separation.
In Fig.~\ref{2PFfixed} (left) the whole setup for wide, intermediate and narrow  shocks is displayed. 
The dark surface represents the radial position of the apparent horizon $z_{\textrm{\tiny AH}}(t,y)$.
The evolution of the energy density of the boundary conformal field theory is shown by a contour  plot located  at the boundary $z=0$. The green lines are geodesics at  various time steps  for a given separation. For narrower shocks the minimum of the apparent horizon is  closer to the boundary and  the influence on  the shape of the geodesics is bigger. 
One can see that the tips of the geodesics tend to avoid the apparent horizon and the evolution of the tips show a similar shape as the apparent horizon.
Once the profile of the geodesics is found the evolution of the 2-point functions can be extracted by computing their length.   
\begin{figure}
\begin{minipage}[t]{0.5\textwidth}
\vspace{-0.8cm}
\hspace{-0.5cm}
\includegraphics[width=\textwidth]{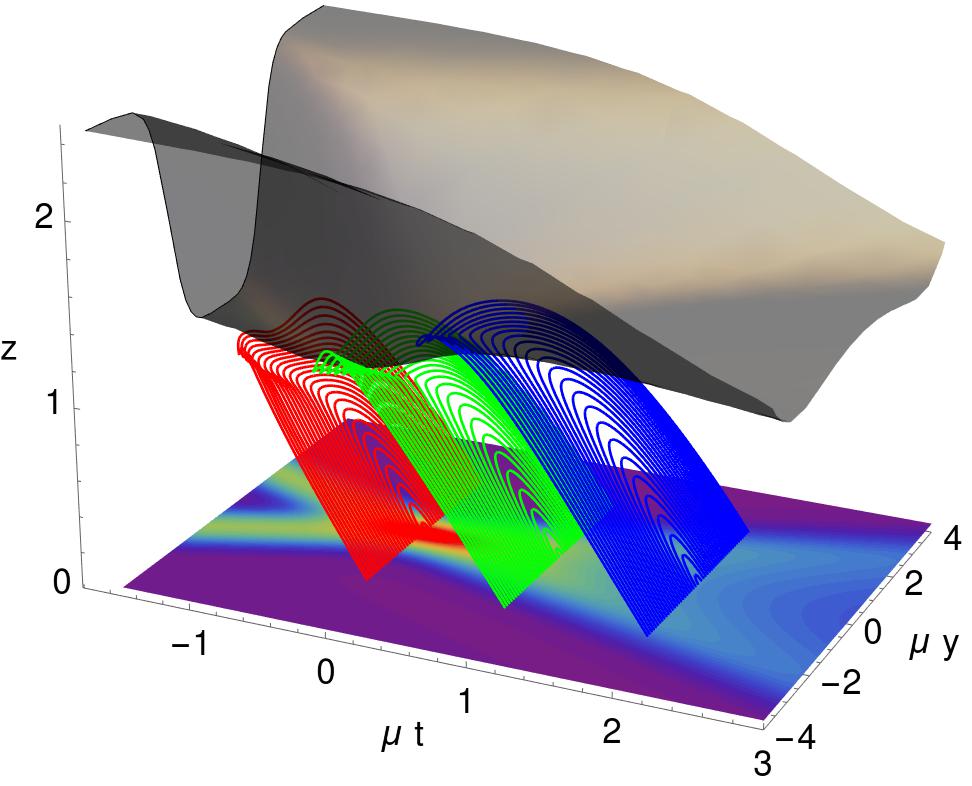}
\end{minipage}
\begin{minipage}[t]{0.5\textwidth}
\vspace{0.1cm}
\includegraphics[width=0.85\textwidth]{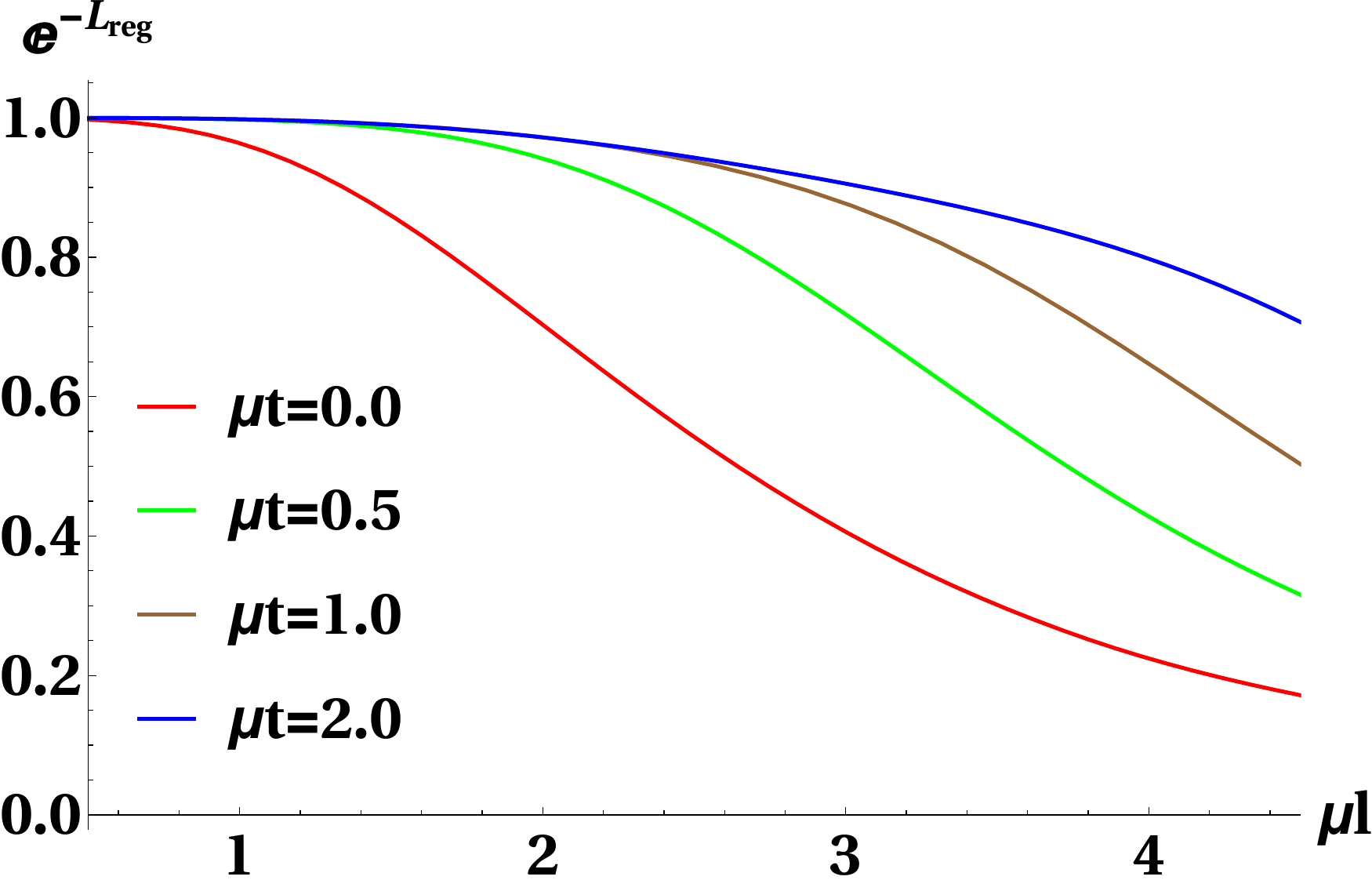}
\end{minipage}
\begin{minipage}[t]{0.5\textwidth}
%\vspace{0.1cm}
\hspace{-0.5cm}
\includegraphics[width=\textwidth]{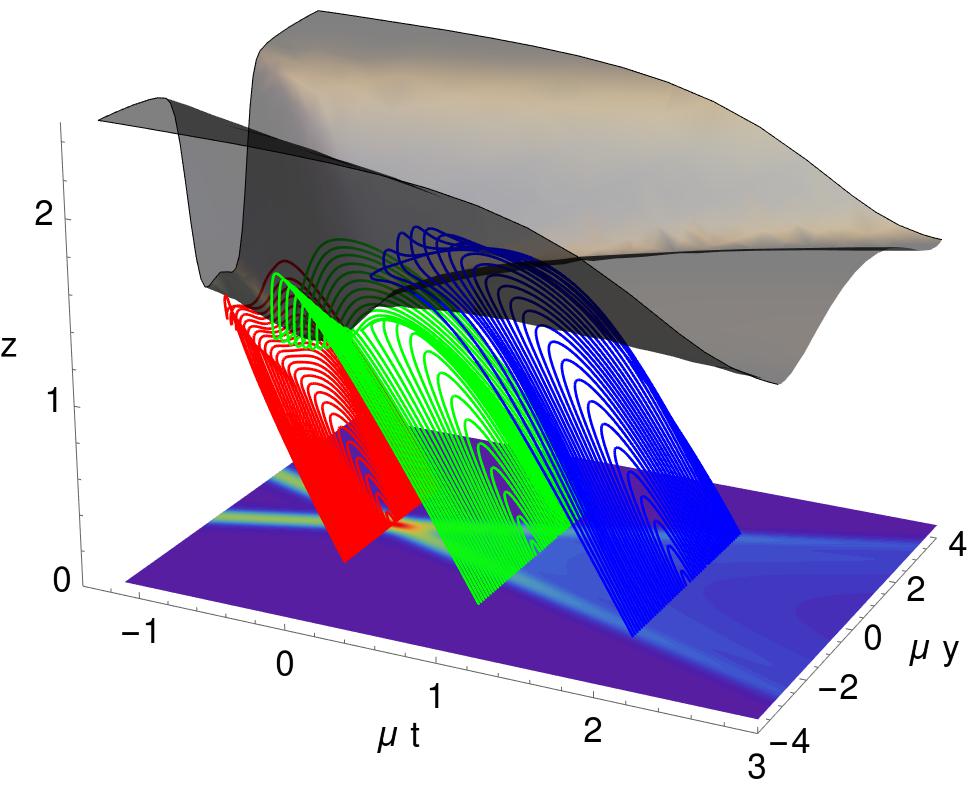}
\end{minipage}
\begin{minipage}[t]{0.5\textwidth}
%\vspace{0.8cm}
\includegraphics[width=0.85\textwidth]{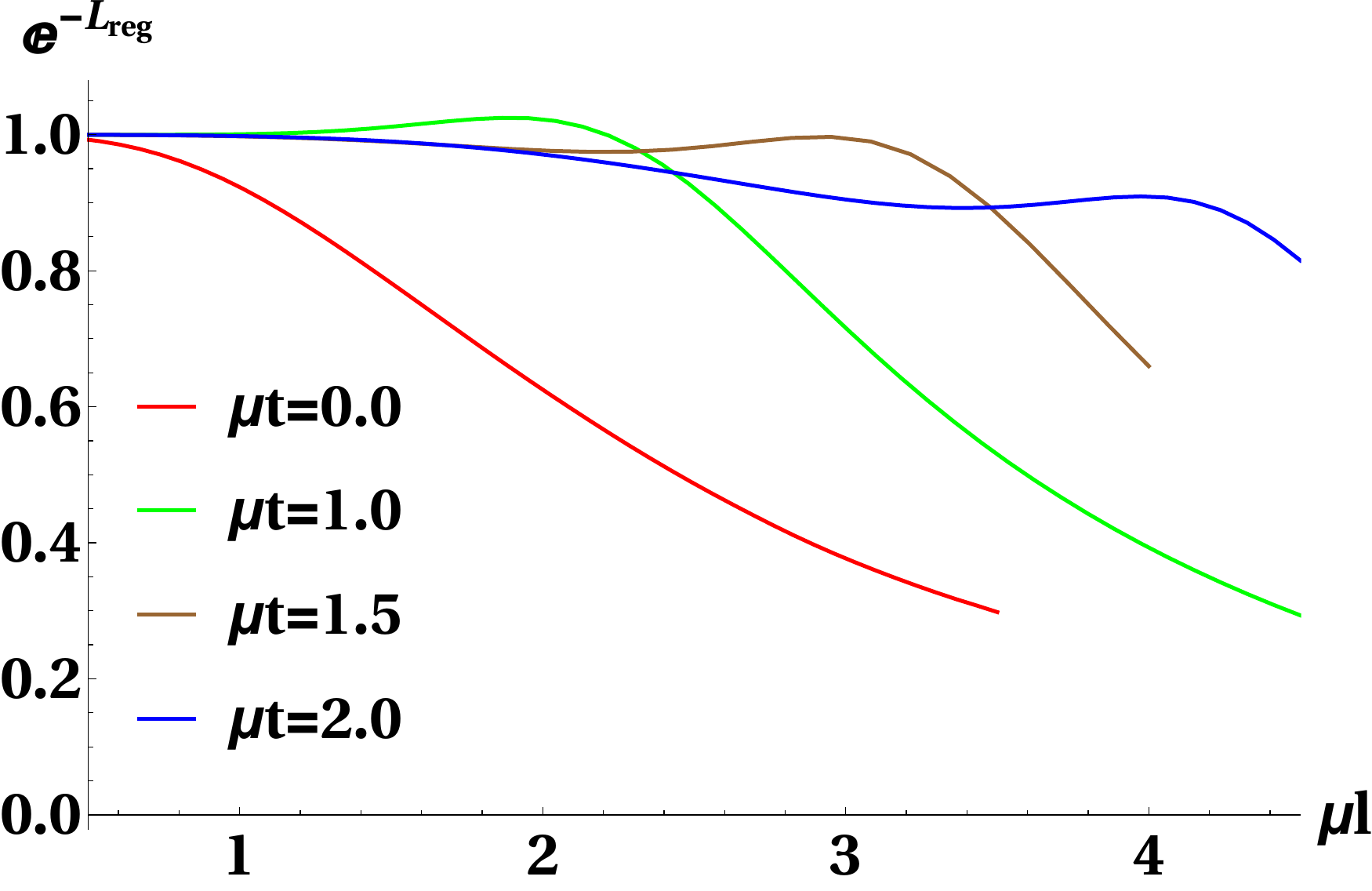}
\end{minipage}
\begin{minipage}[t]{0.5\textwidth}
%\vspace{0.1cm}
\hspace{-0.5cm}
\includegraphics[width=\textwidth]{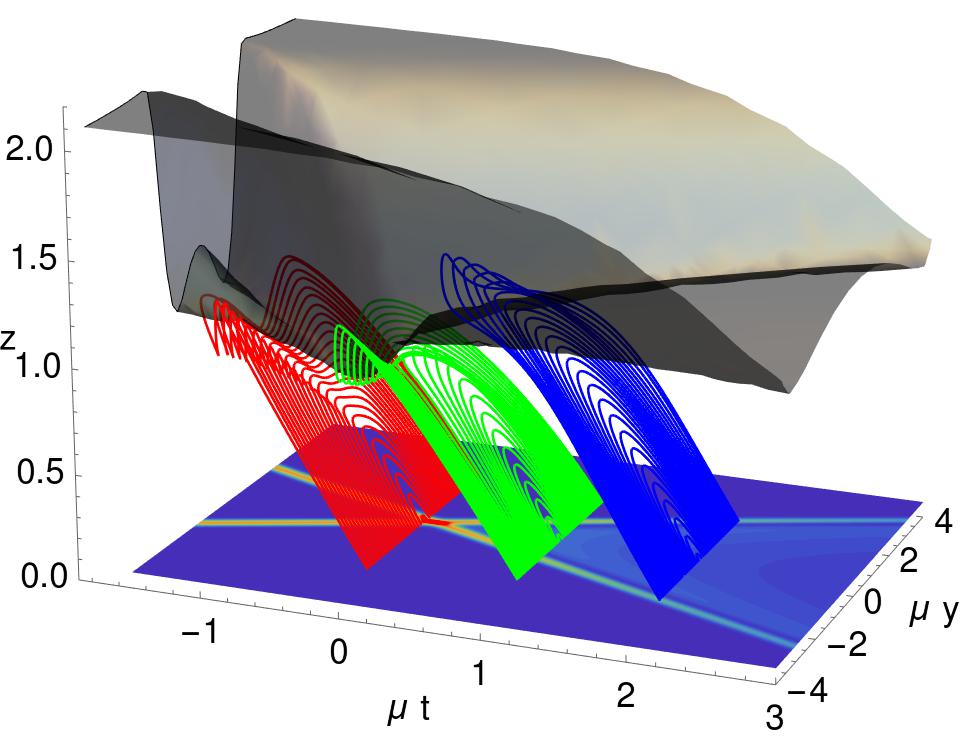}
\end{minipage}
\begin{minipage}[t]{0.5\textwidth}
%\vspace{0.8cm}
\includegraphics[width=0.85\textwidth]{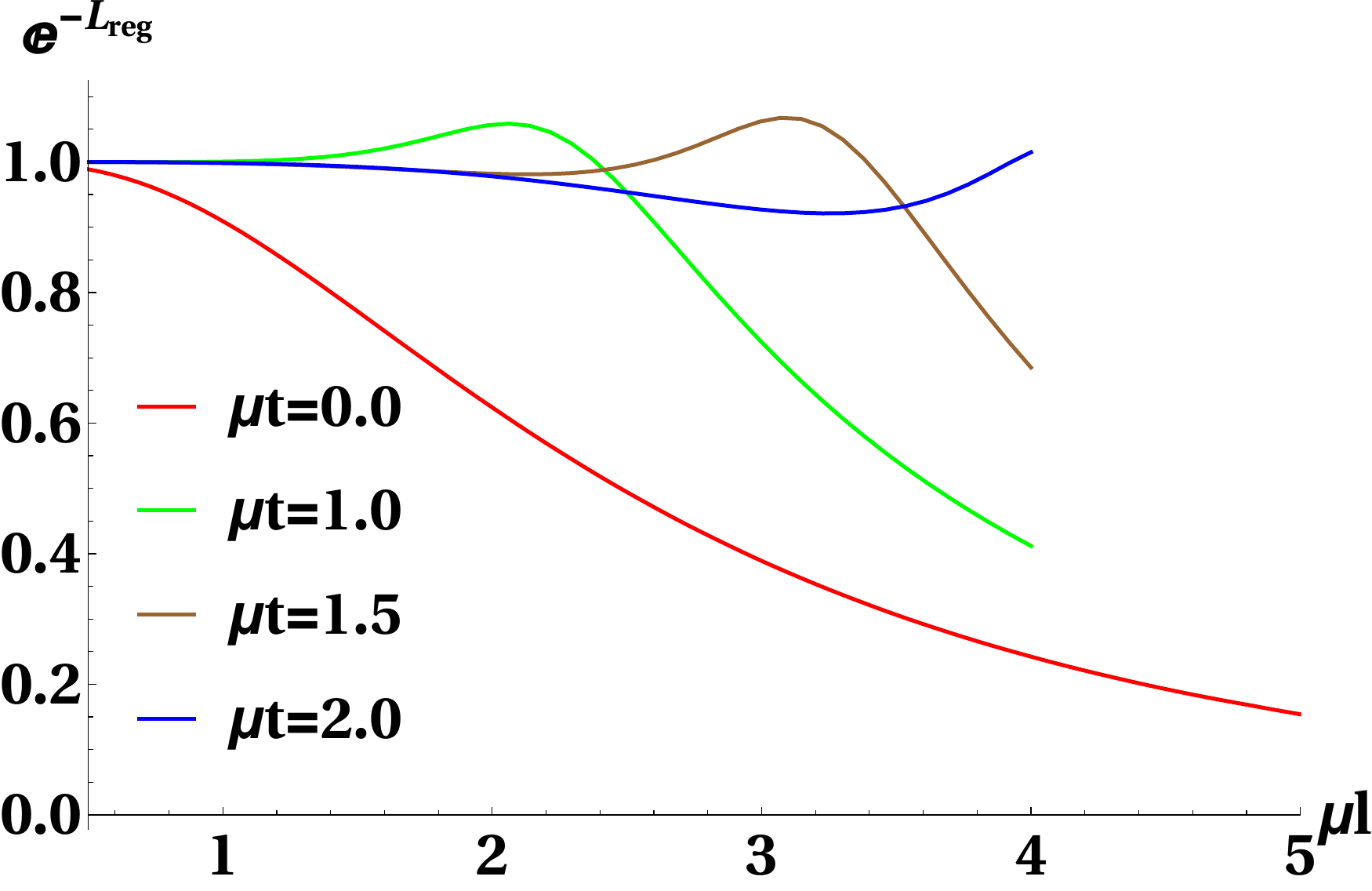}
\end{minipage}
\caption[Geometrical setup and 2-point function (boundary separations).]{Left: Summary of the geometrical setup.
The black surfaces represent the radial position $z_{\textrm{\tiny AH}}(t,y)$ of the apparent horizon; red, green and blue curves are geodesics of various separations at $\mu t=0$, $\mu t=1$ and $\mu t=2$ respectively and at $z=0$ we show a contour plot of the energy density for wide, intermediate and narrow shocks (top to bottom).
Right:  Corresponding evolution of the 2-point function with the boundary separation $\mu l$ at different times.}\label{2PFLevo}
\end{figure}
On the right hand side of Fig.~\ref{2PFfixed} the evolution of the 2-point functions for various boundary separations  for the different  geometries are  displayed.

Let us now summarize the most salient features in the time evolution of the 2-point function during a holographic shock wave collision.
\begin{itemize}
 \item \textbf{rapid onset of linear de-correlation:} 
The system starts in some correlated state. As the shocks are getting closer more and more short range correlations are destroyed and the system rapidly starts to de-correlate in a linear fashion until a local minimum is reached.
The rapid onset of the linear regime is clearly visible for the narrow shocks in Fig.~\ref{2PFfixed}, where for intermediate and wide shocks the onset lies outside our computational domain for larger separations, but the linear regime is still visible. 
 For intermediate and narrow shocks the minimum is located close to $t= 0$ where the energy density is maximal. For wide shocks this minimum is reached significantly before $t=0$. 
\item \textbf{premature de-correlation:} A careful tracking of the position of the minimum as a function of the boundary separation reveals that it is shifted to earlier times as the separation increases. This effect, which is very small and therefore hardly visible in Fig.~\ref{2PFfixed}, is a robust feature of all three kinds of shocks that we have studied.
 \item \textbf{linear correlation restoration:} During the collision, when the shocks interact, new correlations are formed in the system. As the shocks move outwards ($t>0$), the correlations are linearly restored for all three kinds of shocks. 
 \item \textbf{correlation overshooting of narrow shocks:} After the linear restoration regime, the correlations in wide and narrow shocks approach their final values in very different ways.
For intermediate and narrow shocks the correlations significantly overshoot their final values before they finally approach them from above.
In the case of wide shocks this effect is strongly damped and the correlations approach their initial value almost monotonically from below.
\end{itemize}

We switch now to the scaling of the 2-point function with the separation. The holographic setup and the results for the evolution of the 2-point function are displayed in Fig.~\ref{2PFLevo}.
At the collision time ($\mu t=0$) the 2-point function falls off monotonically with the separation in all three cases, although the corresponding geodesics are strongly deformed. For the wide shocks this behavior persists also at later times, where due to the weaker influence of the shocks the correlations fall off more slowly.
For intermediate and narrow shocks an additional maximum appears at $\mu t>0$ which is more pronounced for narrow shocks. The position of this additional maximum is centered around the position of the outgoing shocks.
It is suggestive that narrow shocks which pass through each other almost transparently remain correlated for some time after the collision while wide shocks stop each other before they explode hydrodynamically and the correlations are completely lost.  
This motivated us to study the correlations between the shocks themselves, which we do systematically in Section \ref{app:3}. There we find that the correlations between intermediate and narrow shocks significantly grow after the collision before they start to decay, where the correlations between wide shocks decay immediately. 

\begin{figure}
%\begin{center}
\hspace{-1.cm}
\includegraphics[width=5.5cm]{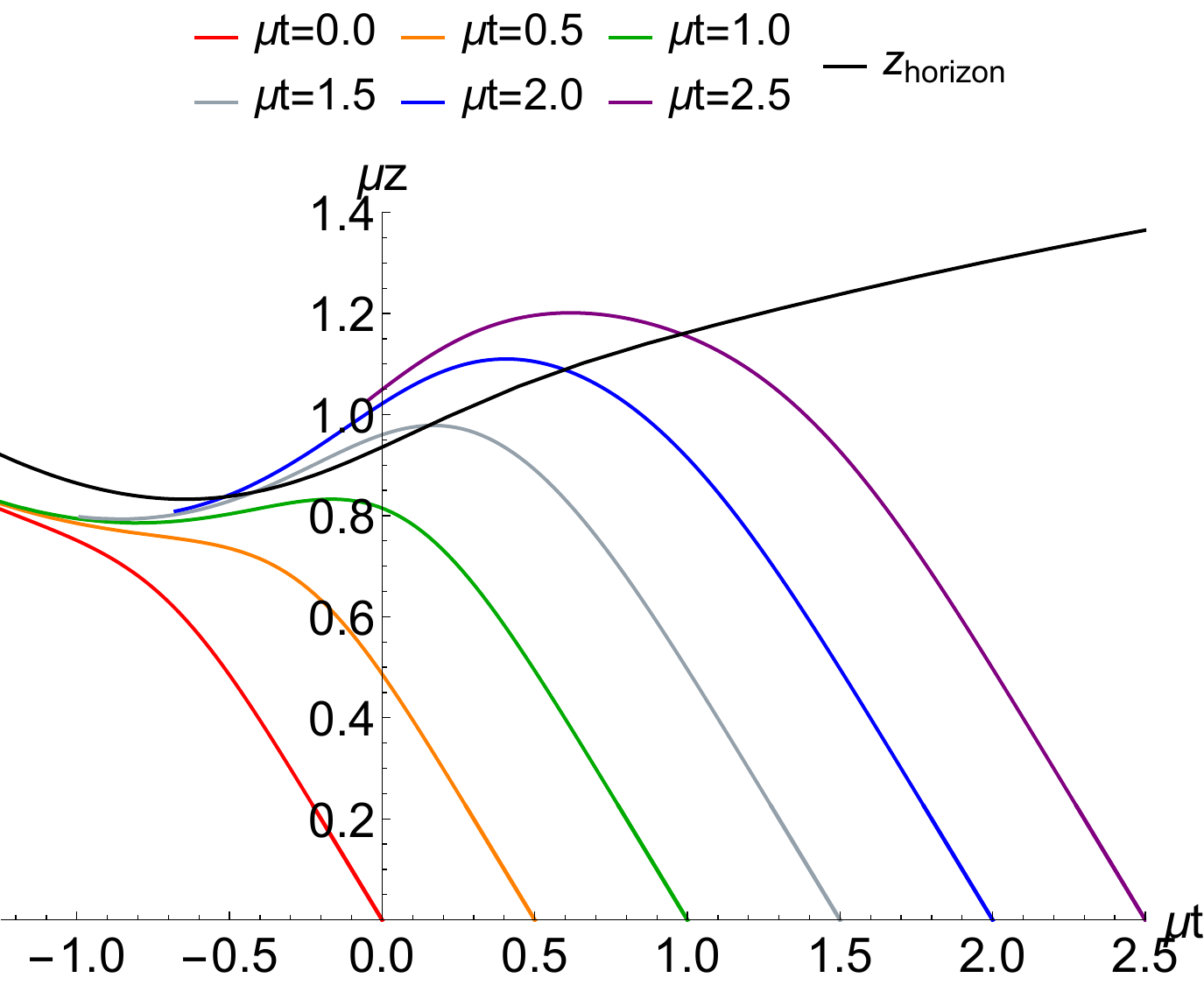}$\;$\includegraphics[width=5.5cm]{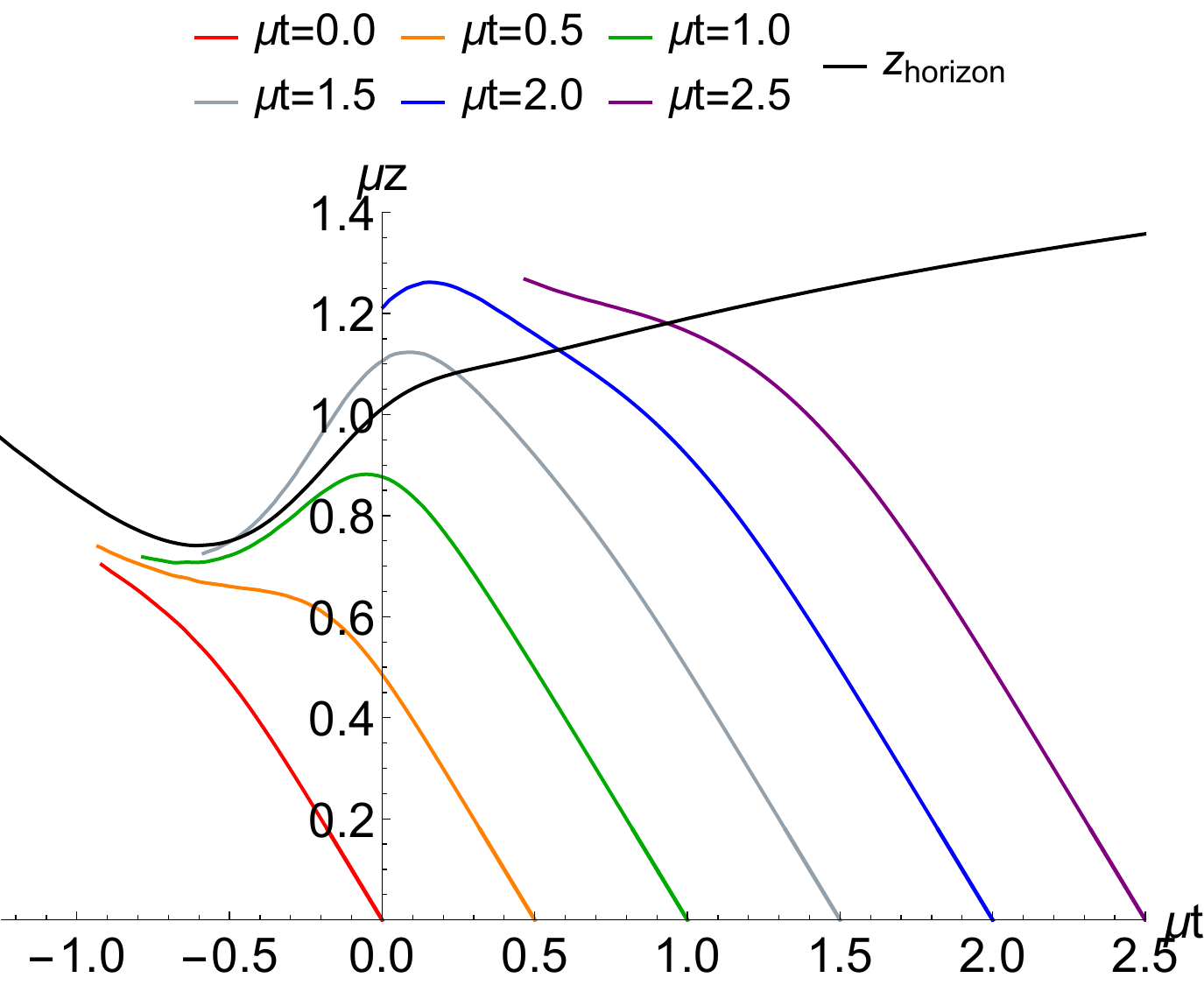}$\;$\includegraphics[width=5.5cm]{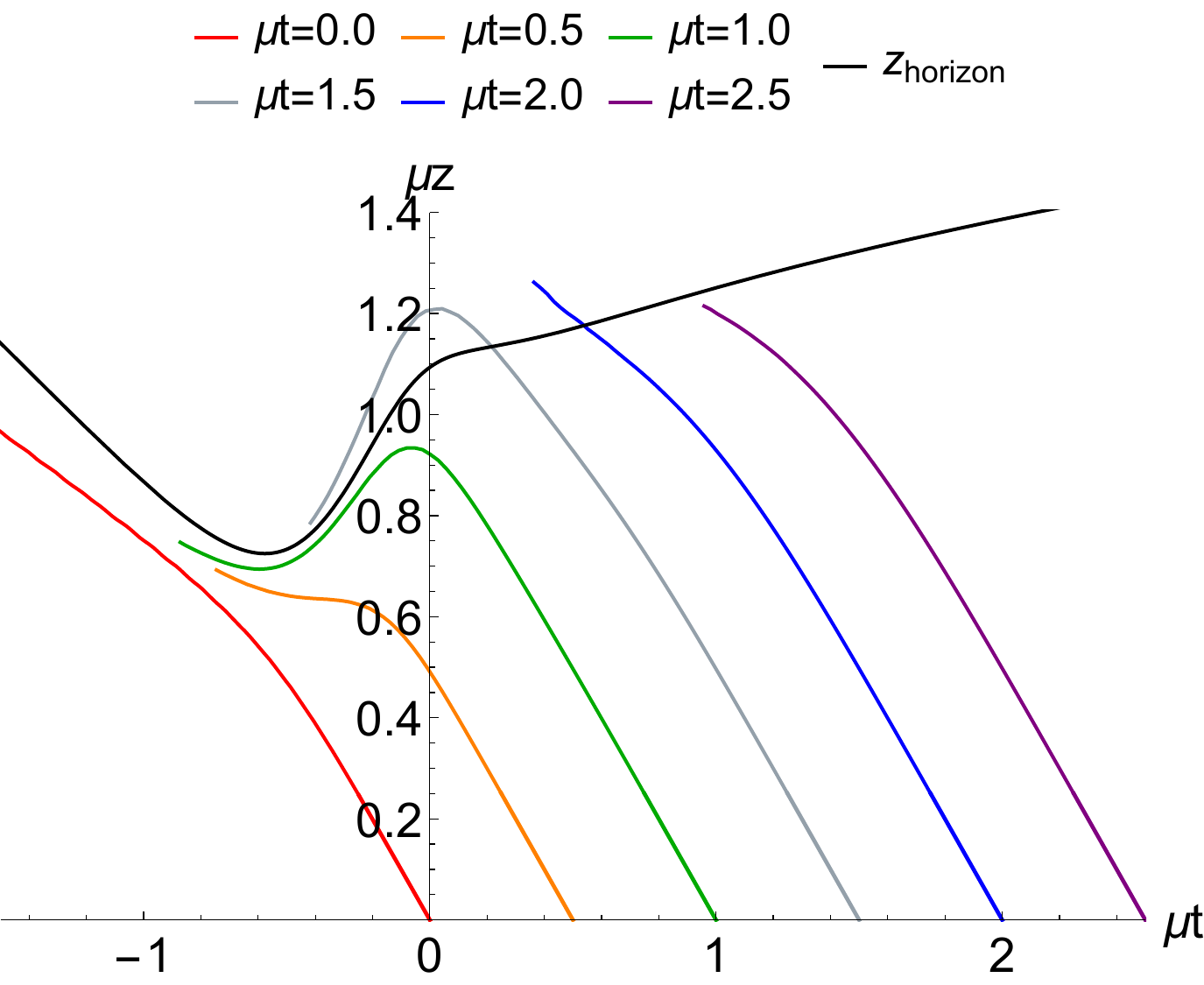}
\caption[Geodesics can extend beyond the apparent horizon.]{\label{GeodesicDip} 
The $z$-position of the geodesics at $y=0$ for several times and separations, starting with $l=0$ near the boundary, and increasing going towards the end of the curve. We show wide, intermediate and narrow shocks (from left to right). The $z$-position of the apparent horizon at $y=0$ is shown in black. At late times and sufficiently large boundary separation in all three cases (wide, intermediate and narrow shocks) geodesics can reach behind the apparent horizon, whereas for early times they reside outside the horizon entirely.}
%\end{center}
\end{figure}

Interestingly, for larger separations the geodesics remain outside the horizon for early times, but they cross the horizon after a time of around $\mu t=1.5$. 
 This can be seen from the blue curves in Fig.~\ref{2PFLevo} and is displayed more clearly in Fig.~\ref{GeodesicDip} where we plot the tip of the geodesic located at $y=0$, for different separations and the position of the apparent horizon at $y=0$.
 This happens for all the initial conditions (wide, intermediate, narrow) we have studied and is in strong contrast to the EE case where we do not find extremal surfaces which cross the horizon.
The crossing after a time of $\mu t=1.5$ is perhaps counterintuitive since geodesics are expected to remain outside the horizon when the system is close to equilibrium. Indeed, hydrodynamics applies after a time $\mu t=0.89$ \cite{Casalderrey-Solana:2013aba}, which is well before the crossing of the geodesics.
At later times presumably the geodesics indeed remain outside again, though our numerics did not allow to determine the precise time at which this is the case. 

%The horizon crossing can be understood by considering the geodesic equation. For this one realizes that at the tip of the horizon both $\partial_y z$ and $\partial_y t$ vanish by the mirror symmetry around $y=0$. It is then the sign of $\partial_y^2 z$ which determines if a geodesic starting horizontally at that point plunges deeper into the bulk, or will go up and finally reach the boundary. The latter geodesics can hence be probed by a boundary two-point function, as is clear from Fig.~\ref{GeodesicDip}. We call the line where $\partial_y^2 z = 0$ the `2PF horizon'\footnote{Strictly speaking it is only an effective horizon for 2PF centered around $y=0$. For non-centered geodesics at the tip $\partial_y t$ does not need to vanish. Nevertheless, prelimenary studies suggest that also those geodesics do not cross the 2PF horizon.}.
 
\subsection{Correlations of colliding shocks}\label{app:3}

 Instead of studying the time evolution of the 2-point function  between  two fixed points  in space, in the context of heavy ion collisions it might be more interesting to actually study the correlation between the two shocks itself. In order to do so, the endpoints of the geodesics follow the maxima of the energy density.
 
\begin{figure}
\begin{minipage}[t]{0.5\textwidth}
\vspace{-0.8cm}
\hspace{-0.5cm}
\includegraphics[width=\textwidth]{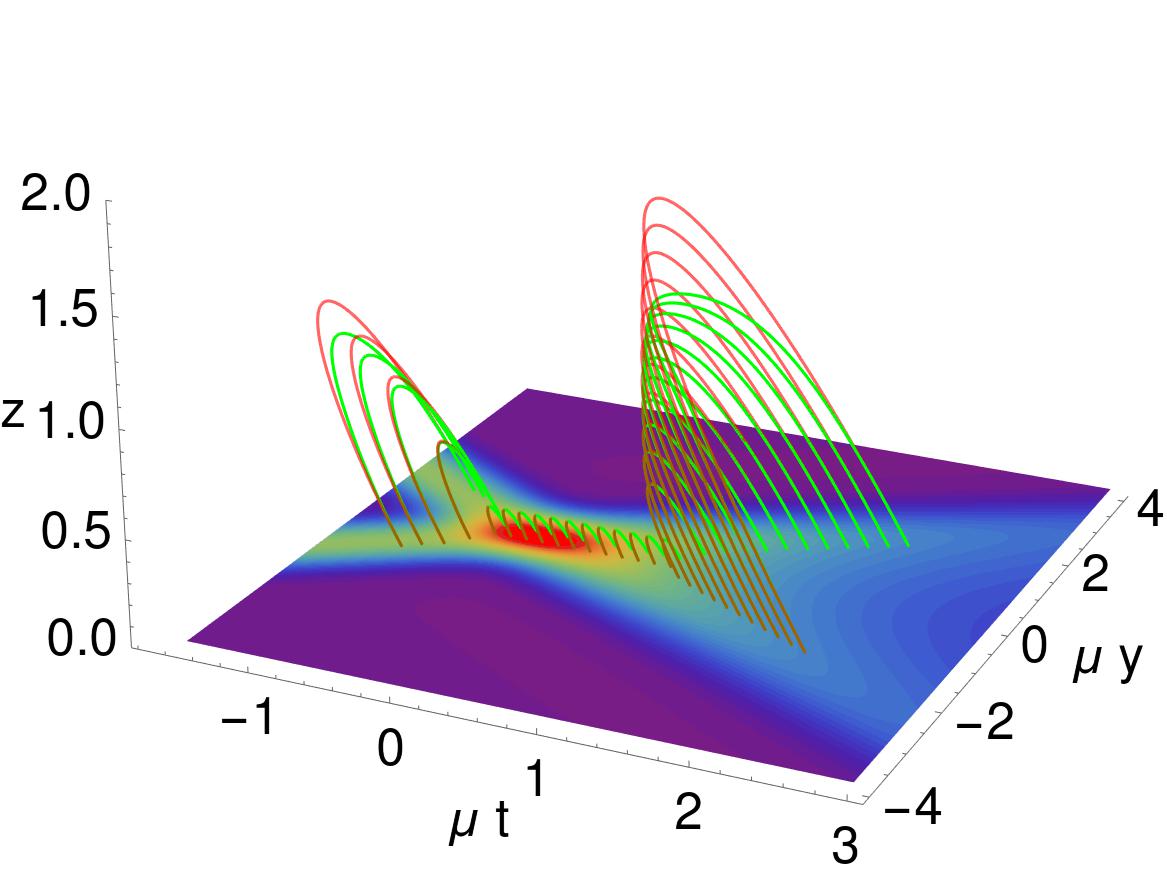}
\end{minipage}
\begin{minipage}[t]{0.5\textwidth}
\vspace{0.1cm}
\includegraphics[width=0.85\textwidth]{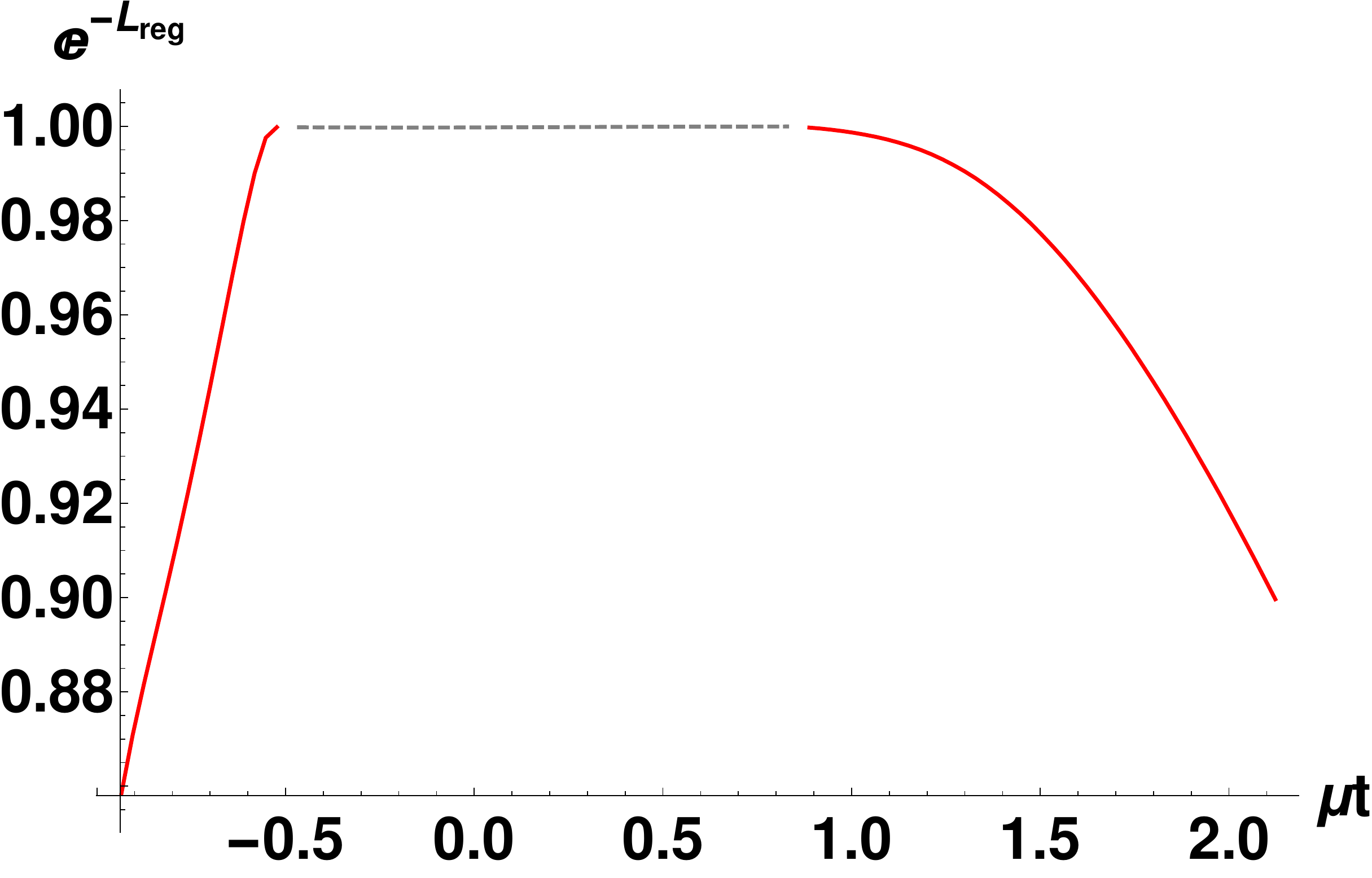}
\end{minipage}
\begin{minipage}[t]{0.5\textwidth}
%\vspace{0.1cm}
\hspace{-0.5cm}
\includegraphics[width=\textwidth]{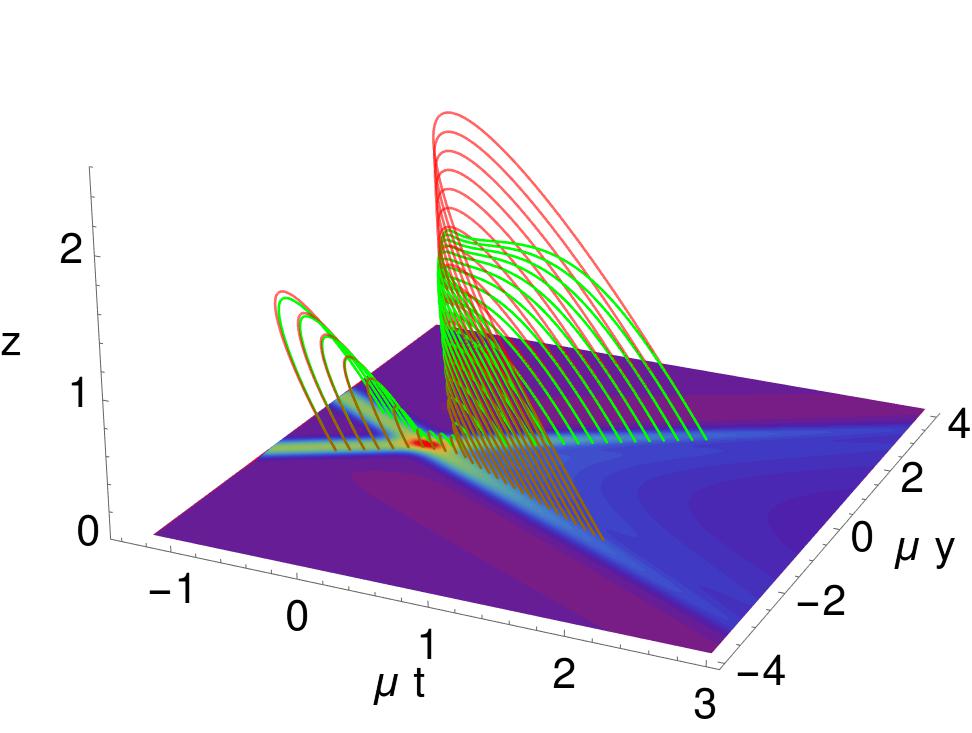}
\end{minipage}
\begin{minipage}[t]{0.5\textwidth}
%\vspace{0.8cm}
\includegraphics[width=0.85\textwidth]{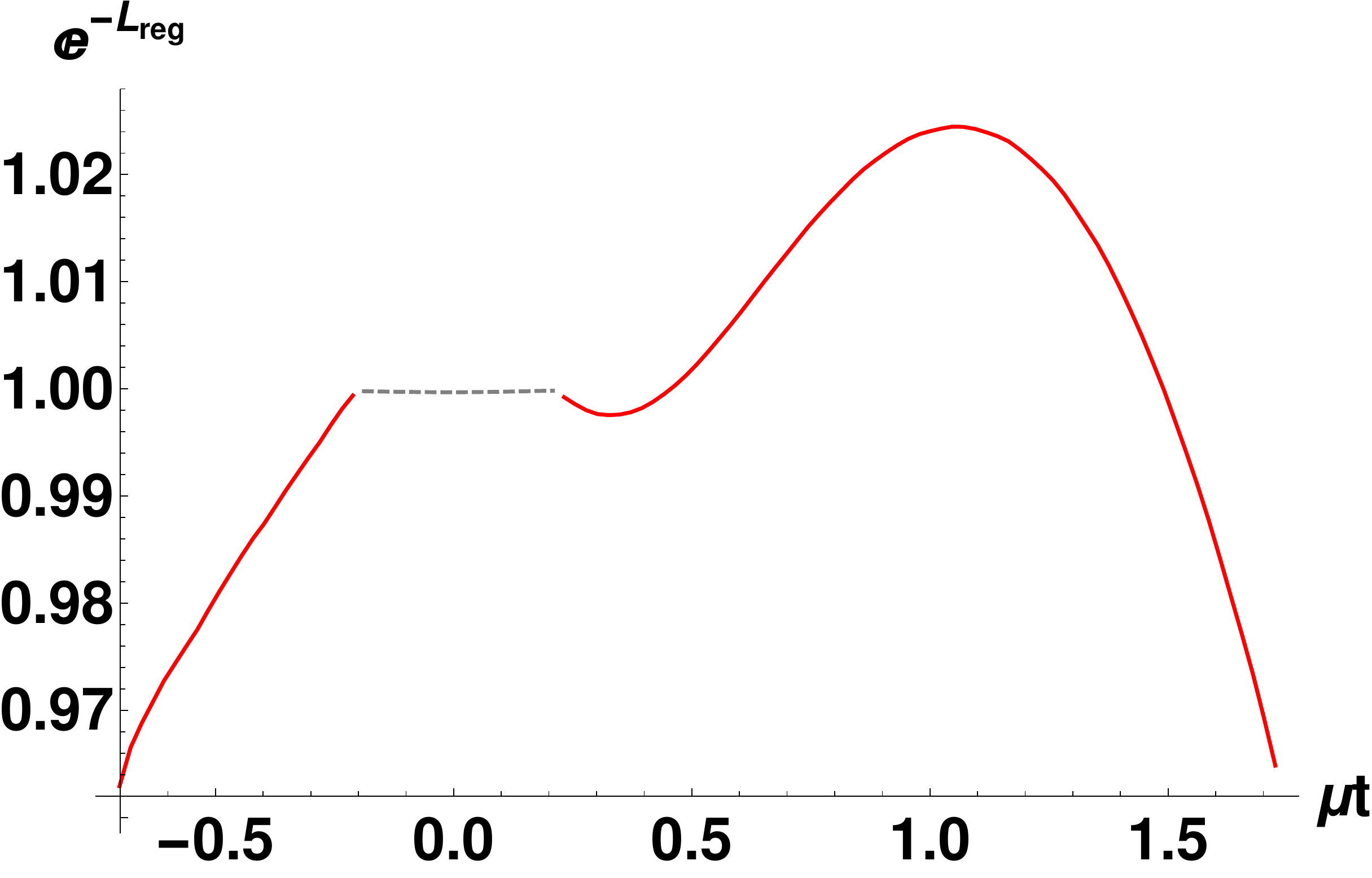}
\end{minipage}
\begin{minipage}[t]{0.5\textwidth}
%\vspace{0.1cm}
\hspace{-0.5cm}
\includegraphics[width=\textwidth]{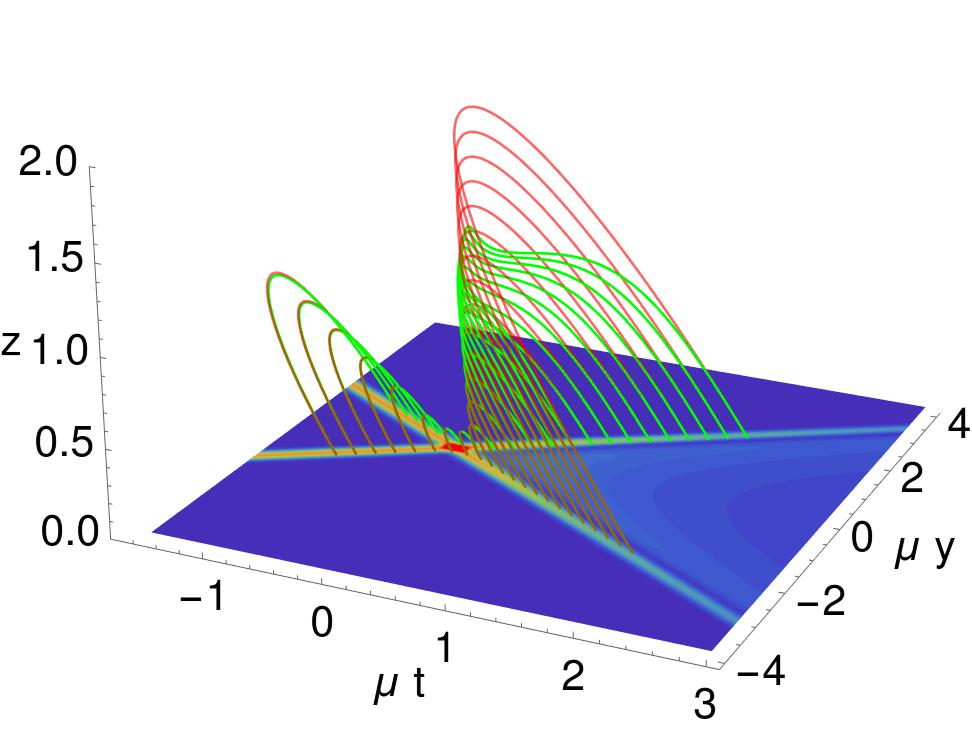}
\end{minipage}
\begin{minipage}[t]{0.5\textwidth}
%\vspace{0.8cm}
\includegraphics[width=0.85\textwidth]{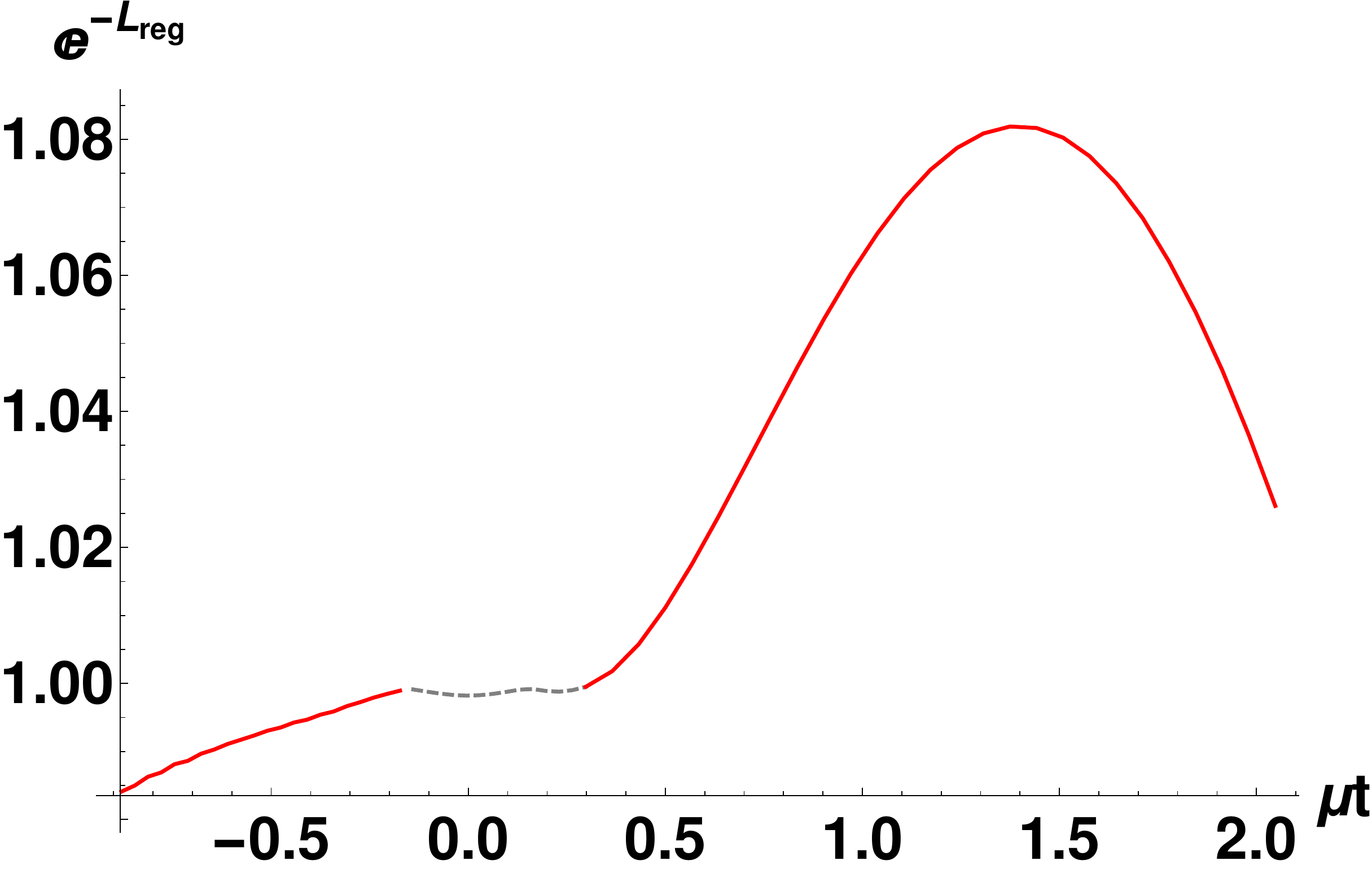}
\end{minipage}
\caption[Correlations of colliding shocks.]{Left: Time evolution of geodesics in the shock wave geometry (green) for wide, intermediate and narrow shocks (top to bottom) pure AdS geodesics (red) with endpoints attached to the position of the maxima in the energy density. Right: Time evolution of the correlation between the shocks; dashed lines indicate the region where only a central maximum in the energy density exist and the separation is fixed to $3*z_{\textrm{\tiny cut}}$.
 }\label{2PFfollow}
\end{figure}

When the separation of the endpoints becomes smaller than three times the cutoff we fix the endpoints to this value until the distance between the two maxima after the collision exceed this value.
  The results are displayed in Fig.~\ref{2PFfollow}, where the geometrical situation is displayed on the left hand side and the time evolution of the 2-point-functions on the right hand side.
 
As already discussed in Section \ref{se:3.2}, for wide shocks the behavior is qualitatively different than for intermediate and narrow shocks. 

As the two wide shocks approach each other their correlation increases almost linearly until it reaches a plateau, which is the point when  the separation of the endpoints is smaller than three times  the cutoff. 
Once the shocks separate again from each other their correlation decreases. 
  
As the shocks get narrower the initial growth slows down  because the shocks start to overlap later. 
After the fixed separation period  a local minimum appears after which the correlations continue to grow to reach another maximum which appears later for narrow shocks. 
  In addition, the maximum correlation is highest for narrow shocks. 

This behavior is reminiscent of the full stopping and transparency scenario for wide and narrow shocks considered in \cite{Casalderrey-Solana:2013aba}. 
 As the wide  shocks start to interact the energy density starts to pile up and all the energy density is contained in a small region after which hydrodynamical explosion occurs. This behavior is also encoded in the 2-point function which reaches a maximum and can only decrease when hydrodynamic explosion occurs. 

For narrower shocks the situation is different. The shocks almost move through each other. Their shape gets altered but no hydrodynamic explosion occurs. The shocks separate from each other and plasma between them forms resulting in a growth of the correlation also after the collision. At sufficiently late times, when the shocks are separated far enough and a hydrodynamical description is applicable the 2-point function decreases rapidly. 

To summarize, there is a general pattern appearing. As the shocks become narrower the initial growth slows down, the maximum  correlation increases and  occurs later.

\section{Entanglement entropy}\label{se:2.3}
In this section we monitor the evolution of EE. 
In time dependent systems the covariant HEE \cite{Hubeny:2007xt} for some boundary region $A$  is obtained  by extremizing the 3-surface functional  
\be\label{area}
{\cal A}=\int \extd^3\sigma\sqrt{\det\Big(\frac{\partial X^{\mu}}{\partial\sigma^a}\frac{\partial X^{\nu}}{\partial\sigma^b}g_{\mu\nu}\Big)}\,
\ee
that ends on the boundary surface $A$.
In the dual field theory the EE is then conjectured to be given by \cite{Ryu:2006bv,Ryu:2006ef,Hubeny:2007xt}
\be
S_{\textrm{\tiny EE}}=\frac{\cal A}{4 G_N}\,.
\ee
Under certain circumstances the problem of finding extremal surfaces can be reduced to finding geodesics in an auxiliary space-time and the problem of solving nonlinear partial differential equation can be circumvented \cite{Ecker:2015kna}. 
In the case at hand this can be achieved by considering a stripe entangling region with finite extent in the longitudinal direction $y$ and infinite extent in the homogeneous transverse directions $(x_1,x_2)$ for which (\ref{area}) simplifies to
\be\label{areastripe}
{\cal A}=\int \extd x_1 \int \extd x_2 \int \extd\sigma\sqrt{\Omega^2h_{\mu\nu}\frac{\partial X^{\mu}}{\partial\sigma}\frac{\partial X^{\nu}}{\partial\sigma}}=V \tilde{L}\,.
\ee
The surface functional (\ref{areastripe}) suffers from two kinds of infinities, one from the integral $V=\int \extd x_1 \int \extd x_2$ over the homogeneous directions and another one from the infinite geodesic length $\tilde{L}$ in the auxiliary spacetime $\Omega^2h_{\mu\nu}$.
Since the infinite volume factor $V$ contains no dynamical information these singularities are avoided by considering EE densities $\frac{S_{\textrm{\tiny EE}}}{V}$.
Analogous to the 2-point function we regularize the geodesic length $\tilde{L}$ by subtracting the corresponding auxiliary vacuum contribution $\tilde{L}_0$.
The observable we compute is the regularized EE density per Killing volume in units of 4$G_N$ 
\be\label{eereg}
S_{\textrm{\tiny reg}}=4G_N\Big(\frac{S_{\textrm{\tiny EE}}}{V}-\frac{S_{\textrm{\tiny EE}}^0}{V_0}\Big)=\tilde{L}-\tilde{L}_0\,.
\ee

%%%%%%%%%%%%%%%%%%%%%%%%%%%%%%%%%%%%%%%%%%%%%%%%%%%%
\subsection{Geodesics in the auxiliary spacetime}\label{se:3.1a}
%%%%%%%%%%%%%%%%%%%%%%%%%%%%%%%%%%%%%%%%%%%%%%%%%%%%
Our aim is to compute the EE for a stripe region with finite extent in $y$-direction and infinite extent in $(x_1,x_2)$ using formula (\ref{eereg}).
Therefore we have to find geodesic lengths $\tilde{L}$ and $\tilde{L}_0$ in the corresponding auxiliary spacetimes.
The auxiliary spacetime, which is related to the metric (\ref{metric}) by a conformal factor $\Omega^2=S^4 e^{2B}$, reads

\be\label{confsubmetric}
\extd\tilde{s}_y^2=S^4 e^{2B}   \big(-A \extd v^2-\frac{2}{z^2}  \extd z \extd v +2 F\extd y\extd v +S^2 e^{-2B} \extd y^2 \big)\,.
\ee
This time we initialize the relaxation algorithm with a geodesic in Poincar\'e patch AdS (\ref{adsmetric}) times a conformal factor $\Omega_0^2=\tfrac{1}{z^4}$
\be\label{confadsmetric}
\extd \tilde{s}_0^2=\frac{1}{z^6}\(-\extd v^2-2 \extd z \extd v + \extd y^2\)\,.
\ee
Like for the Poincar\'e patch AdS geodesics we choose a non-affine parametrization
\ba\label{ansatzEE}
Z_0(\s)&=&Z_{\textrm{\tiny max}}\big(1-\s^2\big)\\
Y_0(\s)&=&\mathrm{sgn}(\sigma)\Big(-\frac{l}{2} + \frac{W Z_0(\sigma)^4}{4}\, {}_2F_1 \lk \tfrac{1}{2},\tfrac{2}{3},\tfrac{5}{3}; W^2 Z_0(\sigma)^6\rk\Big)\\
V_0(\s)&=&t-Z_0(\s)
\ea
where $W\!=\!\frac{\pi^\frac{3}{2} \Gamma\lk 5/3\rk^3}{8 l^3 \Gamma\lk 7/6\rk^3}$ ensures that the two branches, discriminated by $\mathrm{sgn}(\s)$, join smoothly at $Z_{\textrm{\tiny max}}\!=\!\frac{2l}{\sqrt{\pi}}\Gamma(\tfrac{7}{6})\Big/\Gamma(\tfrac{5}{3})$. 
The affine parameter $\tau$ in terms of $\sigma$ reads
\be
\tau(\sigma)=\frac{\mathrm{sgn}(\s)}{2Z_{\textrm{\tiny max}}^2(1-\sigma^2)}\, {}_2F_1\lk \tfrac{1}{2},-\tfrac{1}{3},\tfrac{2}{3};W^2 Z_{\textrm{\tiny max}}^{12}(1-\sigma^2)^6\rk
\ee
and the Jacobian evaluates to
\be\label{JacobianEE}
J(\sigma)=\frac{\extd^2\tau}{\extd\sigma^2}\Big/\frac{\extd\tau}{\extd\sigma}=\frac{-51\s+145\s^3-205\s^5+159\s^7-65\s^9+11\s^{11}}{(2-\s^2)(1-\s^2)(3-3\s^2+\s^4)(1-\s^2+\s^4)}\,.
\ee
Using the ansatz (\ref{ansatzEE}) and the corresponding Jacobian (\ref{JacobianEE}) in the relaxation algorithm allows us to compute geodesics in the auxiliary spacetime (\ref{confsubmetric}).

The bulk parts of the geodesic lengths in Eq.~(\ref{eereg}), which are the contributions from $z>z_{\textrm{\tiny cut}}$, follow from integrating the line elements (\ref{confsubmetric}) and (\ref{confadsmetric}) 
\bse
\ba
\tilde{L}^{\textrm{\tiny bulk}}&=&\int_{\s_-}^{\s_+} \extd\s S^2 e^{B}\sqrt{-A \dot{V}^2-\frac{2}{Z^2}\dot{Z}\dot{V}+2F\dot{V}\dot{Y} +S^2 e^{-2B}\dot{Y}^2},\\
\tilde{L}_0^{\textrm{\tiny bulk}}&=&\int_{\s_-}^{\s_+} \extd\s \frac{1}{Z_0^3}\sqrt{- \dot{V_0}^2-2 \dot{Z}_0 \dot{V}_0 + \dot{Y}_0^2},
\ea
\ese
where in this case the bounds of the integral $\sigma_\pm$, implementing the IR-cutoff at $z\!=\!z_{\textrm{\tiny cut}}$, are given by
\be
\s_\pm=\pm\sqrt{1-\frac{z_{\textrm{\tiny cut}}}{Z_{\textrm{\tiny max}}}}\,.
\ee
We build the near boundary part ($0\le z \le z_{\textrm{\tiny cut}}$), like for the 2-point function, from the asymptotic solution of the geodesic equation in the conformal spacetime, which
leads to the following near-boundary expansion
\bse\label{asympSolutionEE}
\ba
 Z(z) & = & z \\
 V(z) & = & t_0-z+v_4 z^4+\frac{a_4 z^5}{5}+O\left(z^6\right) \\
 Y(z) & = & \frac{l}{2}+y_4 z^4+\frac{f_4 z^5}{5}+O\left(z^6\right) \\
 J(z) & = & \frac{3}{z}+\left(2 a_4-4 b_4\right) z^3+O\left(z^6\right), \\
\ea
\ese
where the normalizable modes $a_4(v,y)$, $b_4(v,y)$ and $f_4(v,y)$ are evaluated at $v=t_0$ and $y=\pm\tfrac{l}{2}$.  We again have two undetermined constants $v_4$ and $y_4$, which now appear two orders higher than for the case of the 2-point function. Again we also have the analytic solution in the auxiliary pure AdS space time
\ba\label{asympAnsatzEE}
Z_0(z)&=&z\\
V_0(z)&=&t-Z_0(z)\\
Y_0(z)&=&\pm\Big(-\frac{l}{2} + \frac{W Z_0(z)^4}{4}\, {}_2F_1 \lk \tfrac{1}{2},\tfrac{2}{3},\tfrac{5}{3}; W^2 Z_0(z)^6\rk\Big)\nonumber\\
      &=&\pm\big(-\frac{l}{2}+\frac{W}{4}z^4\big)+\mathcal{O}(z^{10})\\
J_0(z)&=&\frac{3-6W^2z^6}{z-W^2z^7}=\frac{3}{z}-3W^2z^5+\mathcal{O}(z^{11}).
\ea
The near boundary contribution to the geodesic length  for both endpoints evaluates to
\bse
\ba
\tilde{L}^{\textrm{\tiny bdry}}-\tilde{L}_0^{\textrm{\tiny bdry}}&=& \big(b_4-\frac{a_4}{2}\big)z+\big(\partial _tb_4-\frac{7 \partial _ta_4}{20}\big)z^2\nonumber\\
                   &+& \frac{1}{120} (20 \partial _y\partial _tf_4-13 \partial _t^2a_4+70 \partial _t^2b_4+7 \partial _y^2a_4+2 \partial _y^2b_4+960 y_4^2-960 t_4^2)z^3\nonumber\\
                   &+& \mathcal{O}(z^{4}),
\ea
\ese
where the divergent term cancels out again. Now this formula is clearly more useful, as the two leading contributions do not depend on the unknown coefficients $v_4$ and $y_4$, which hence allows to reduce the cut-off dependence significantly.
The regularized EE of Eq.~(\ref{eereg}) is the sum of the bulk contribution and the near boundary contribution   
\be\label{Sreg}
S_{\textrm{\tiny reg}}=(\tilde{L}^{\textrm{\tiny bulk}}-\tilde{L}_0^{\textrm{\tiny bulk}})+(\tilde{L}^{\textrm{\tiny bdry}}-\tilde{L}_0^{\textrm{\tiny bdry}})\;.
\ee
As for the 2-point function we checked the convergence of $S_{\textrm{\tiny reg}}$ with the gridsize in the range from 50 up to 400 gridpoints and find again that for more than $200$ gridpoints the change in $S_{\textrm{\tiny reg}}$ is smaller than $\mathcal{O}(10^{-5})$ which is the same order as the allowed residual we choose in the relaxation algorithm.

To achieve cutoff independence of $S_{\textrm{\tiny reg}}$ turns out to be more delicate than for the 2-point function. Now for a range $z_{\textrm{\tiny cut}}=[0.05,0.1]$ we obtain a slightly worse cutoff dependence of $\mathcal{O}(10^{-3})$ which is however sufficient for our qualitative studies where $S_{\textrm{\tiny reg}}=\mathcal{O}(10^{-1})$ and the influence of the cutoff can be estimated to be $\approx 1 \%$ (see Appendix \ref{app:2}).
Again we choose $200$ gridpoints to discretize our geodesics and set $z_{\textrm{\tiny cut}}=0.075$ in all the calculations we present below.
%%%%%%%%%%%%%%%%%%%%%%%%%%%%%%%%%%%%%%
\subsection{Evolution of entanglement entropy}
%%%%%%%%%%%%%%%%%%%%%%%%%%%%%%%%%%%%%%

%%%%%%%%%%%%%%%%%%%%%%%%%%%%%%%%%%%%%%%%%%%
\begin{figure}
\begin{center}
\hspace*{-0.0cm}\includegraphics[scale=.25]{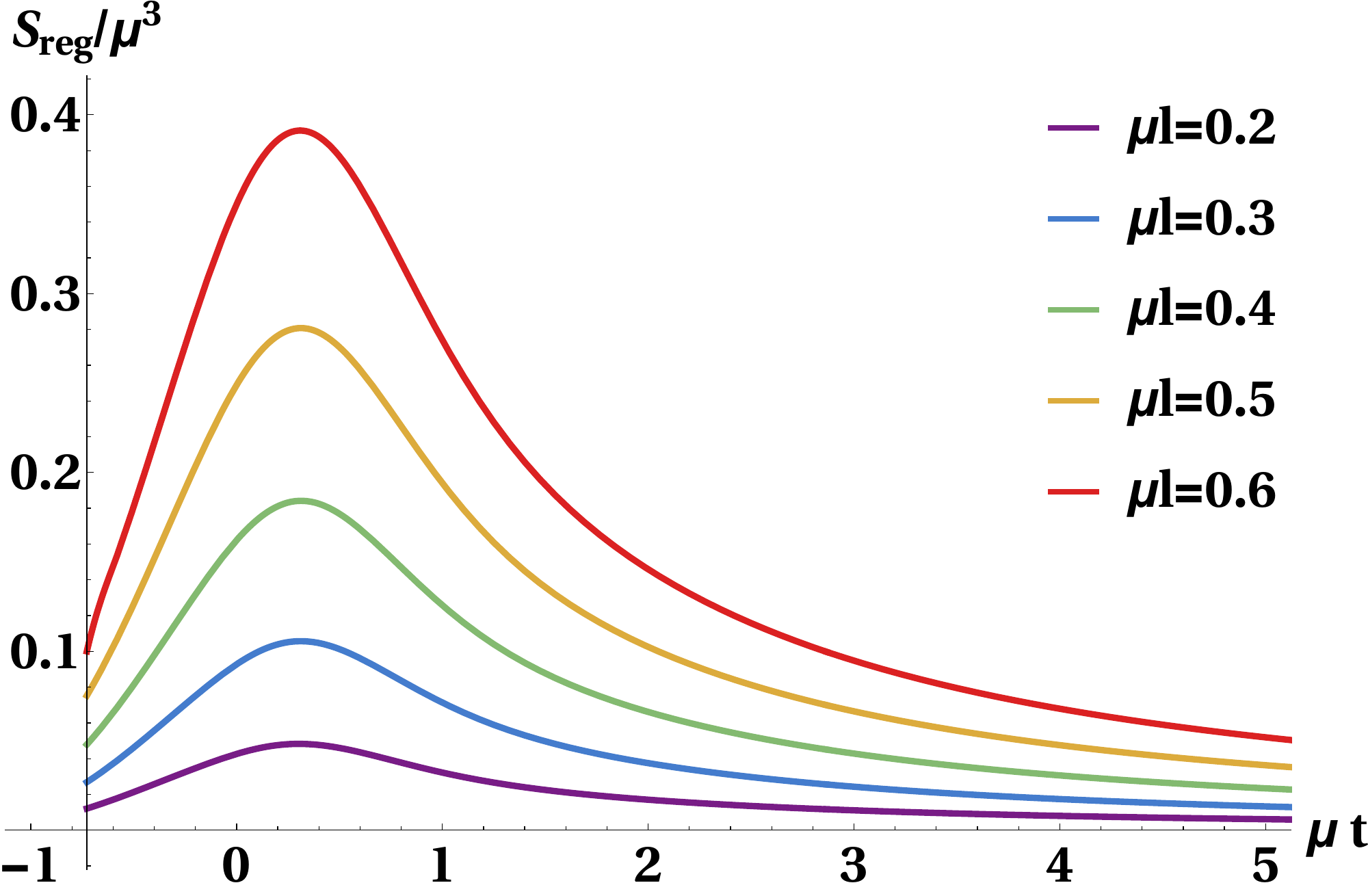}$\;\;\;$\includegraphics[scale=.25]{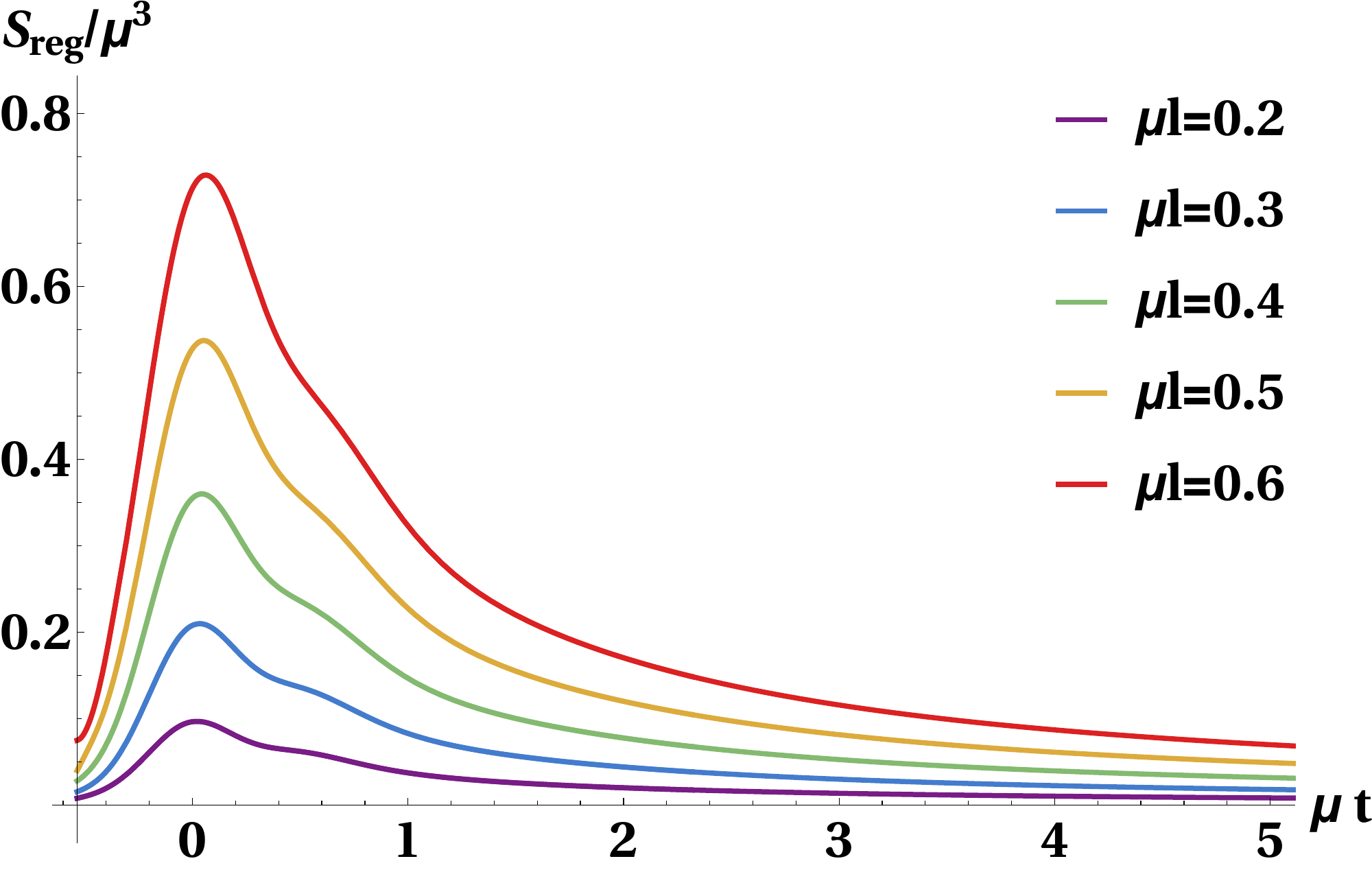}$\;\;\;$\includegraphics[scale=.25]{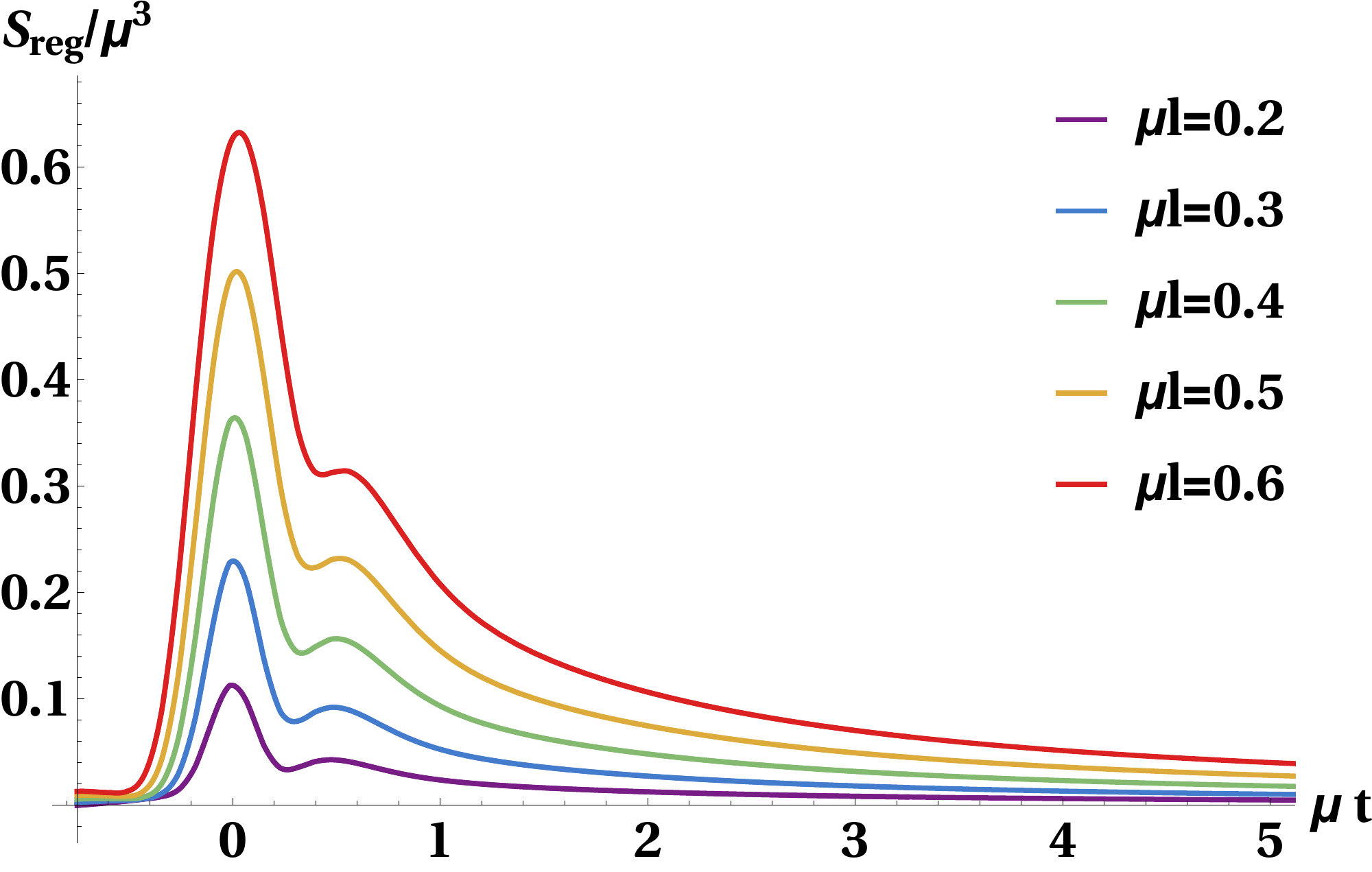}
\caption[Time evolution of entanglement entropy.]{\label{EEevolution} 
Evolution of the EE for different separations of  width $\mu l$ of the stripe region for wide  (left), intermediate  (middle) and narrow  (right) shocks. 
 }
\end{center}
\end{figure}
%%%%%%%%%%%%%%%%%%%%%%%%%%%%%%%%%%%%%%

In this section we present our numerical results for the EE.
The shape of EE as a function of time originates from a complicated interplay between the different metric functions appearing in the energy momentum tensor. 
However, most features can be understood in terms of  energy density and pressures. 
In Fig.~\ref{EEevolution} we display the time evolution of HEE for various separations in the two different scenarios. 
It can be characterized by four distinct regions:
\begin{enumerate}
\item \textbf{rapid initial growth}: 
Once some energy density enters the entangling region the rapid initial growth starts.
The narrower the shocks the more rapidly the initial growth happens, because the rate at which the energy density enters the entangling region is bigger than for wider shocks. 
\item \textbf{linear growth}: The linear growth starts when the two shocks start to overlap and the energy piles up, with a steeper slope for larger separations. This is the same behavior as the post-local equilibration growth after a global quench \cite{Liu:2013iza}.
The maximum occurs with a short delay compared to the maximum energy deposited in the entangling region, with a more pronounced  delay for wider shocks.
\item \textbf{post collisional regime}: The post collisional regime is quite different for the three cases considered. 
For wide shocks the EE falls off without any additional features. 
In the case of intermediate shocks a small shoulder appears. In the case of narrow shocks this shoulder turns into a new feature, where an additional minimum appears and the EE starts growing again until a second maximum is reached. 
The minimum happens approximately at a time when the longitudinal pressure becomes negative. The existence or absence of a minimum of EE in this regime thus serves as an order parameter to discriminate between narrow and wide shocks. 
\item \textbf{late time regime}: At late times we find a polynomial fall off behavior
\be\label{lateEE}
S_{\textrm{\tiny reg}}\approx a_{w,i,n} (\mu t)^{-b_{w,i,n}}\;,
\ee
where the coefficient  $a_{w,i,n}$ depends on the initial conditions and the separation.
In Table \ref{fit} we give the late time behavior extracted from the time interval $\mu t=[2,6]$ for different separations.
The late time behavior can be compared to the late time behavior of an effective entropy density 
\be
s_{\textrm{\tiny eff}}(t)=\int\limits_{-l/2}^{l/2} \mathrm{d}y\, S^3(r_h,t,y)\;,
\ee
where the function $S$ is evaluated at the position of the apparent horizon and integrated over the same intervals as for the EE. 
The late time behavior is displayed in Table \ref{fit1} and barely depends on the separation. 
It is expected on general grounds that at very late times and large separations, far beyond our computational domain,  the effective entropy density and EE show the same fall off behavior. 
\end{enumerate}

\begin{figure}
\begin{minipage}[t]{0.5\textwidth}
\vspace{-0.8cm}
\hspace{-0.5cm}
\includegraphics[width=\textwidth]{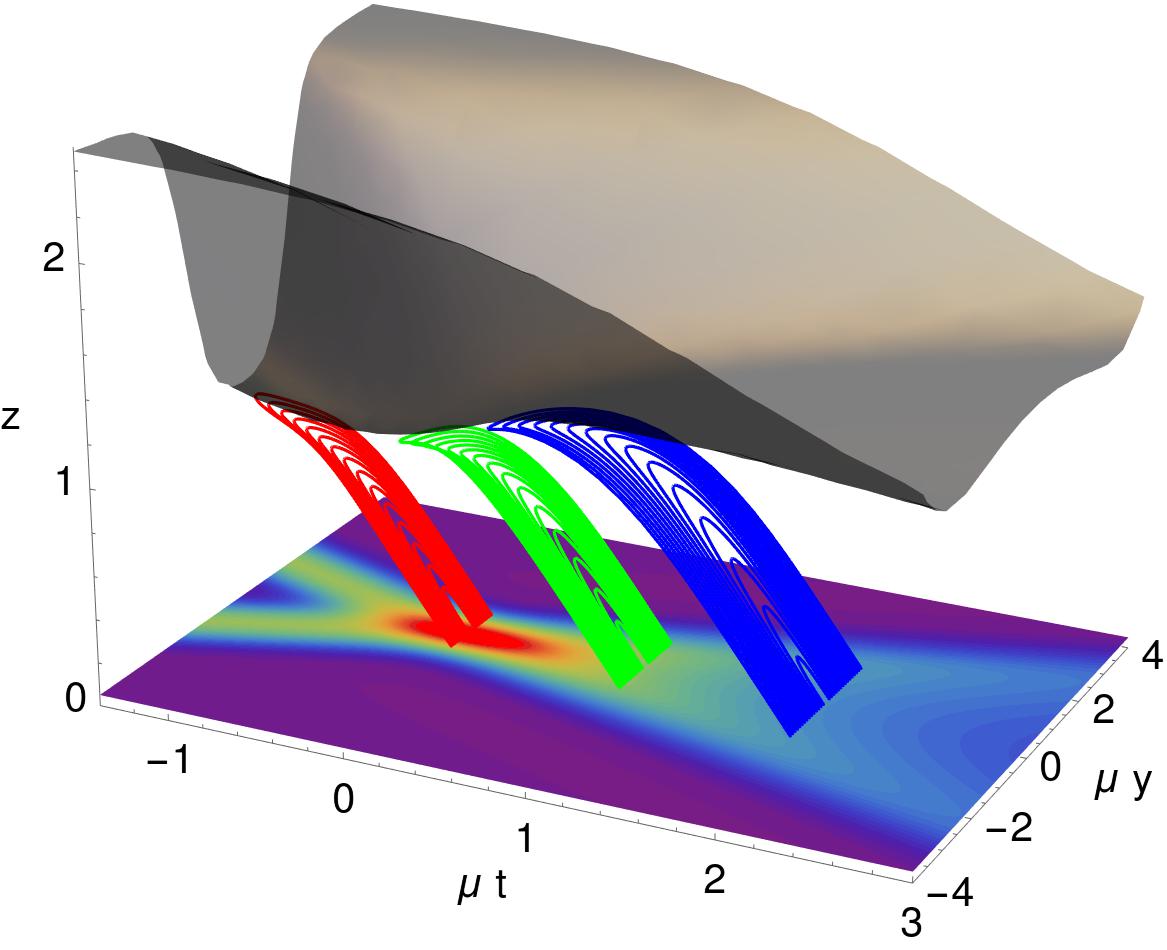}
\end{minipage}
\begin{minipage}[t]{0.5\textwidth}
\vspace{0.1cm}
\includegraphics[width=0.85\textwidth]{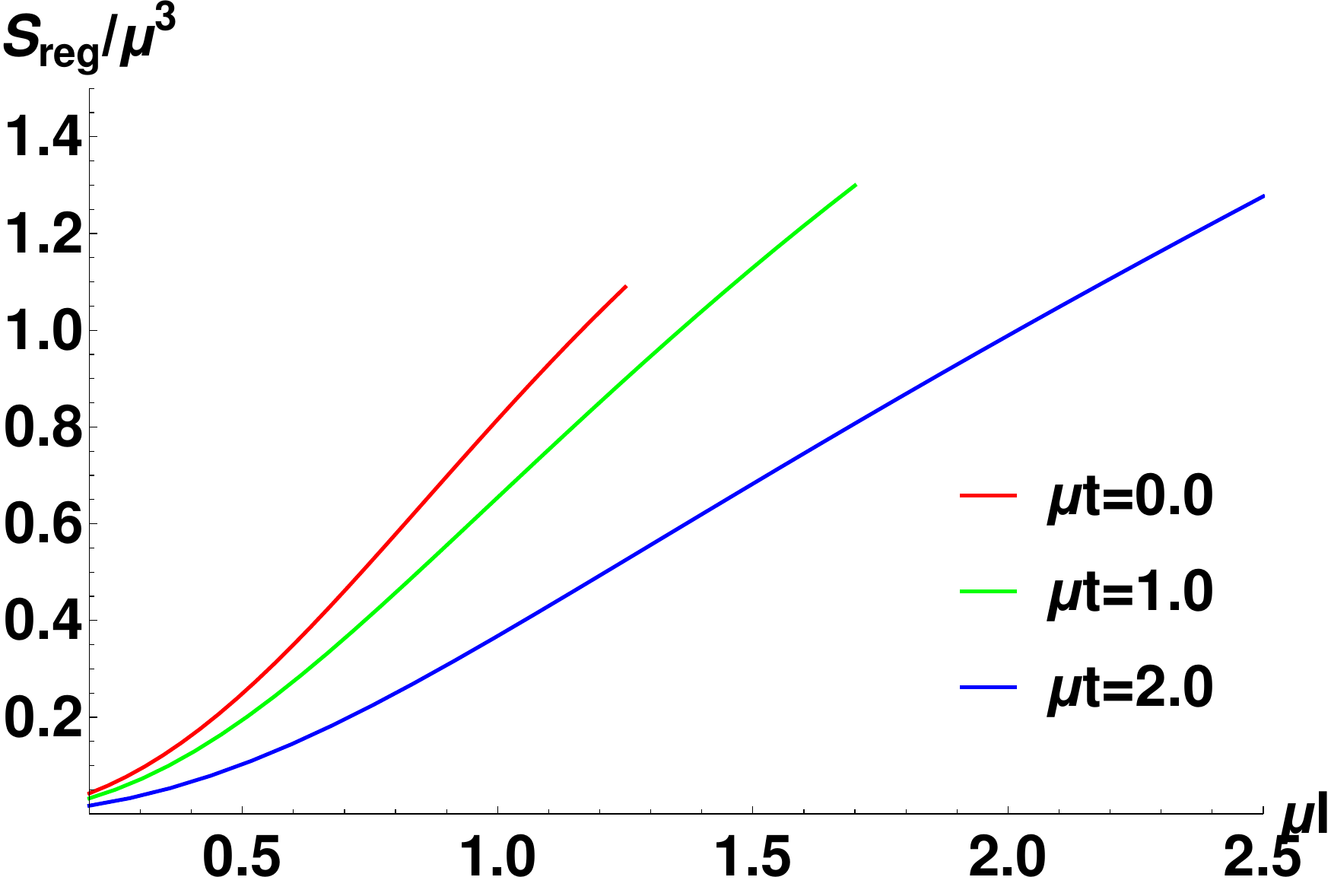}
\end{minipage}
\begin{minipage}[t]{0.5\textwidth}
%\vspace{0.1cm}
\hspace{-0.5cm}
\includegraphics[width=\textwidth]{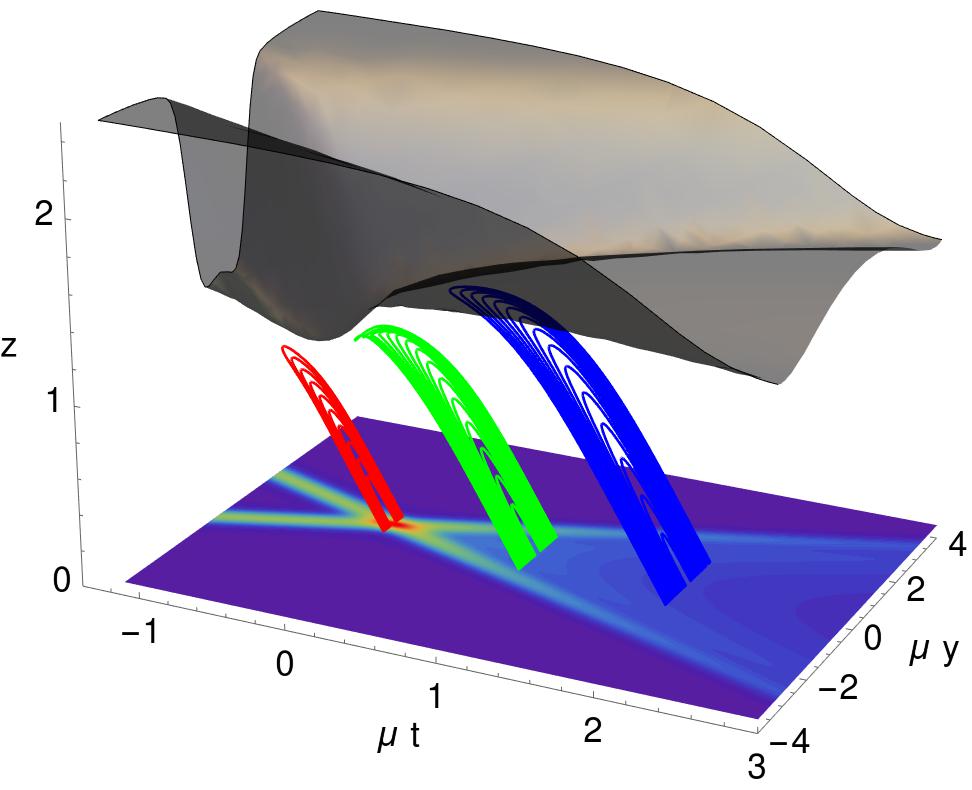}
\end{minipage}
\begin{minipage}[t]{0.5\textwidth}
%\vspace{0.8cm}
\includegraphics[width=0.85\textwidth]{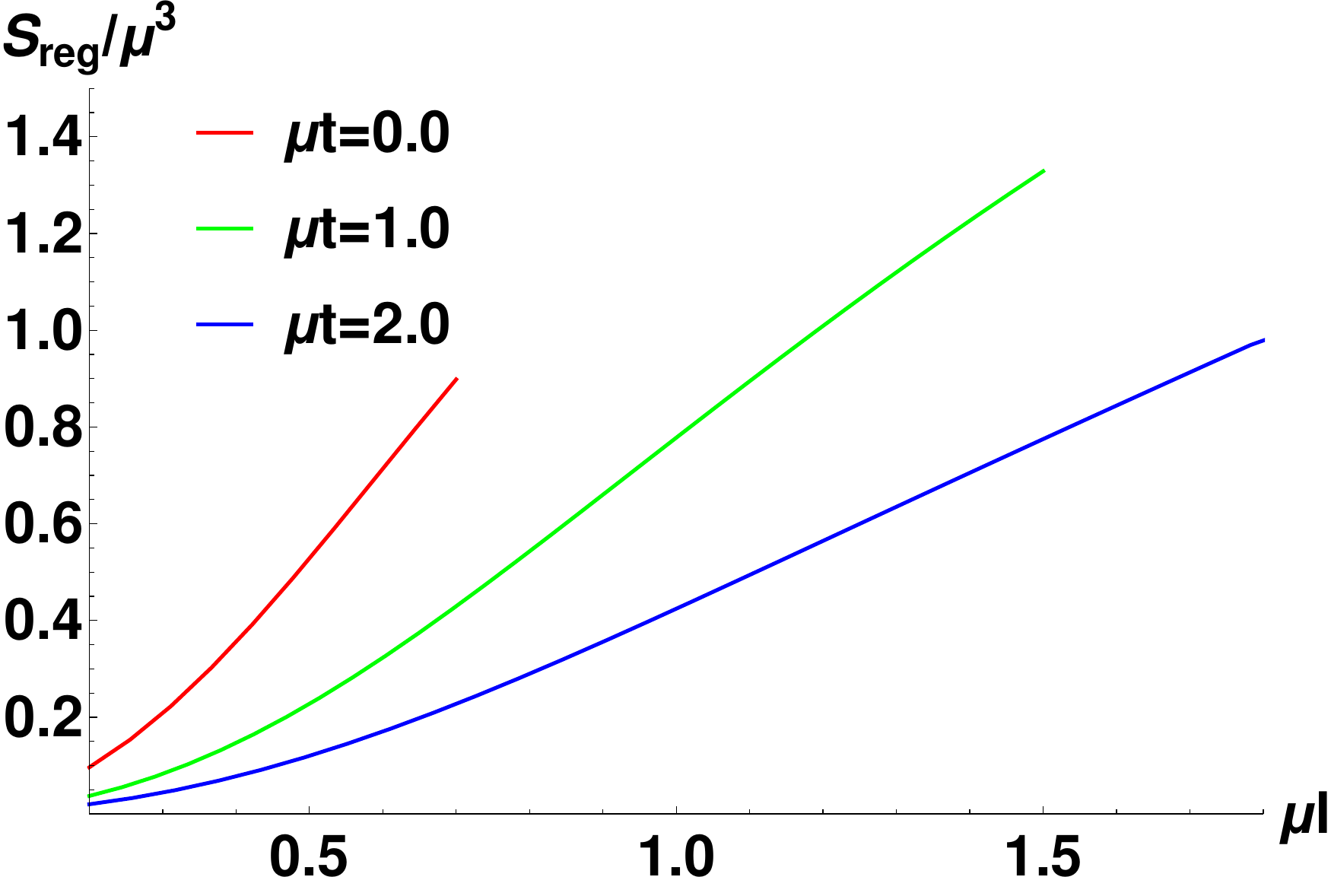}
\end{minipage}
\begin{minipage}[t]{0.5\textwidth}
%\vspace{0.1cm}
\hspace{-0.5cm}
\includegraphics[width=\textwidth]{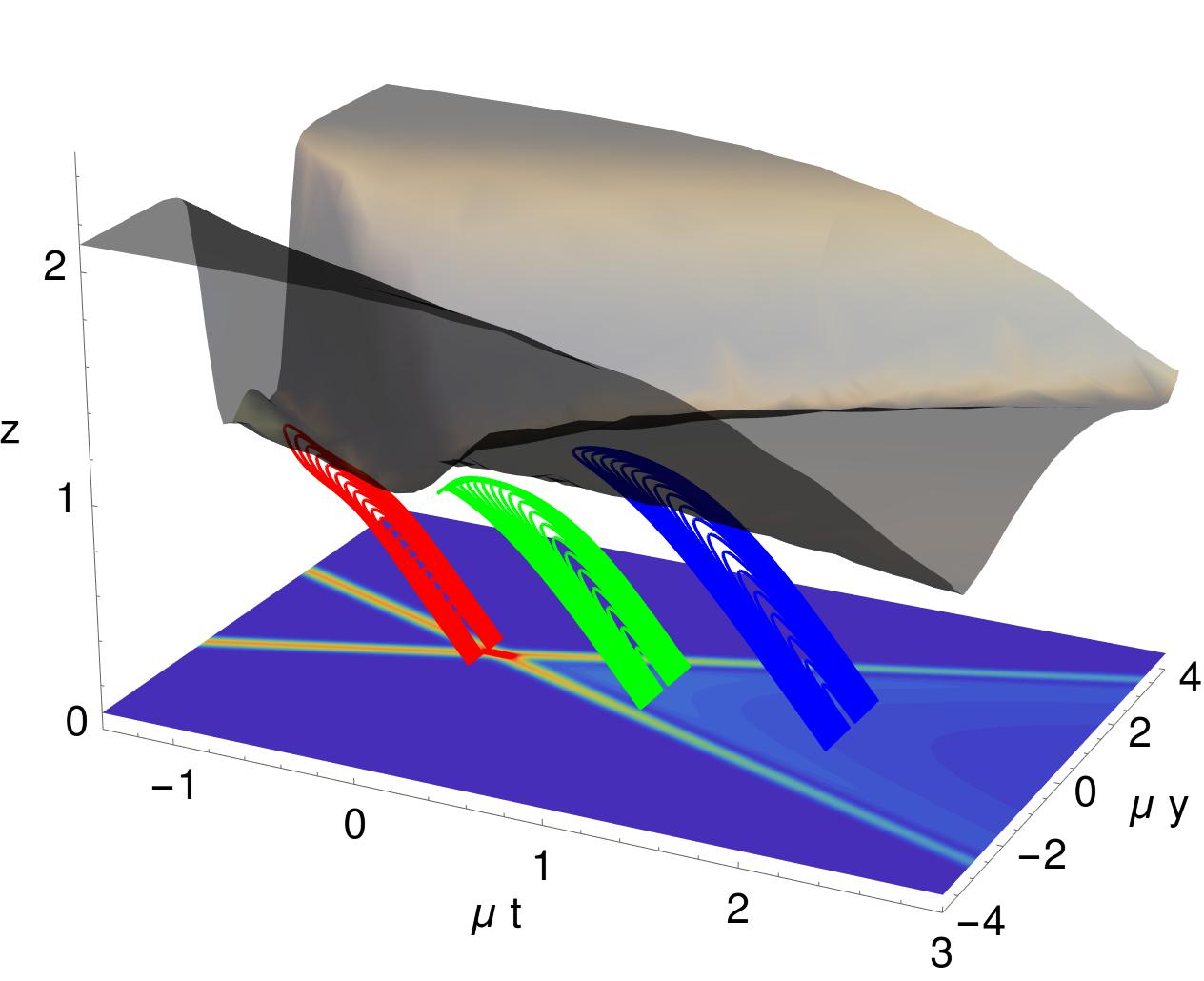}
\end{minipage}
\begin{minipage}[t]{0.5\textwidth}
%\vspace{0.8cm}
\includegraphics[width=0.85\textwidth]{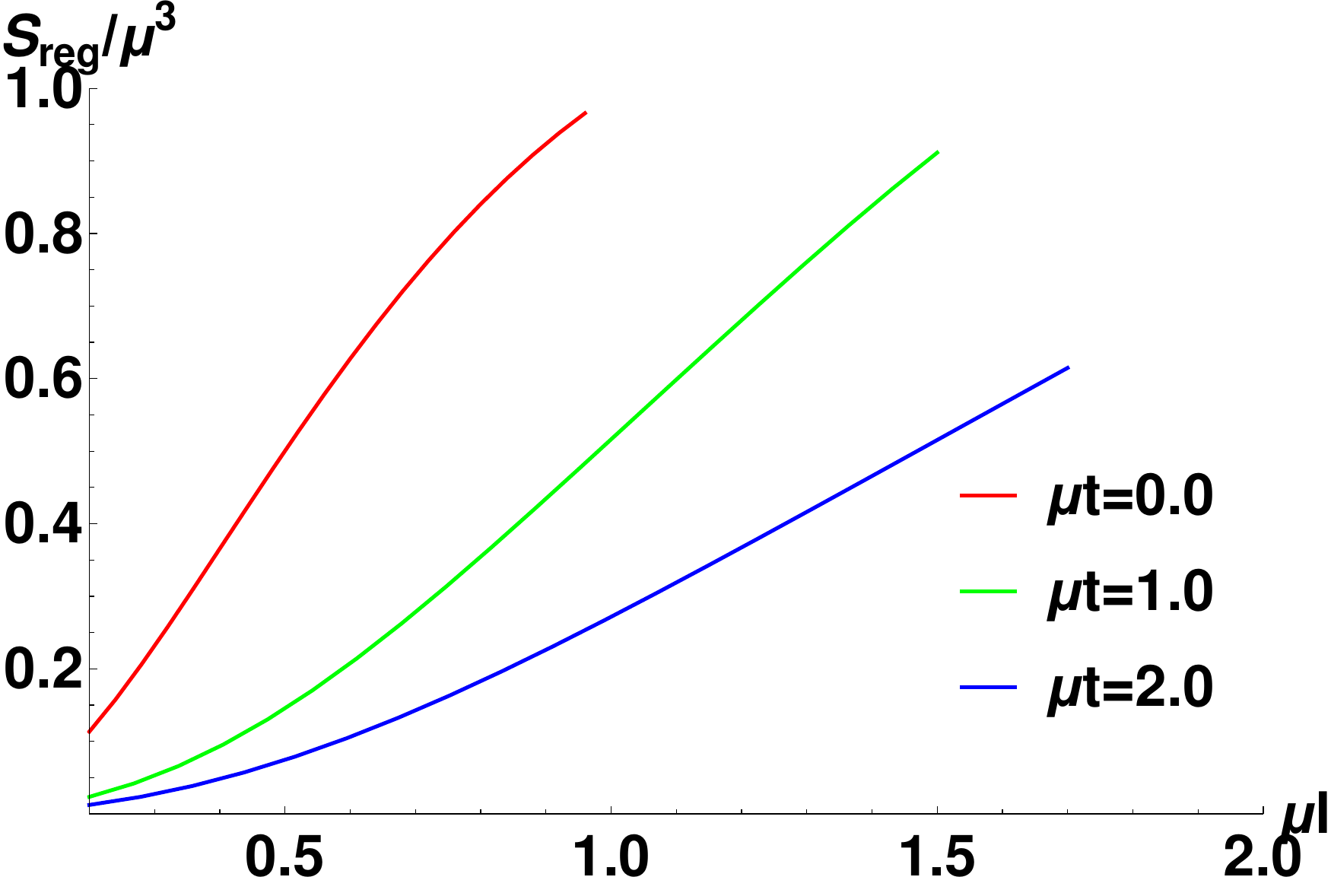}
\end{minipage}
\caption[Geometrical setup and entanglement entropy (boundary separations).]{Left: Summary of the geometrical setup.
The black surfaces represent the radial position $z_{\textrm{\tiny AH}}(t,y)$ of the apparent horizon; red, green and blue curves are geodesics of various separations at $\mu t=0$, $\mu t=1$ and $\mu t=2$ respectively and at $z=0$ we show a contour plot of the energy density for wide, intermediate and narrow shocks (top to bottom).
Right:  Corresponding evolution of the EE with the boundary separation $\mu l$ at different times.}\label{EELevo}
\end{figure}

\begin{figure}
%\begin{center}
\hspace{-1.cm}
\includegraphics[width=5.5cm]{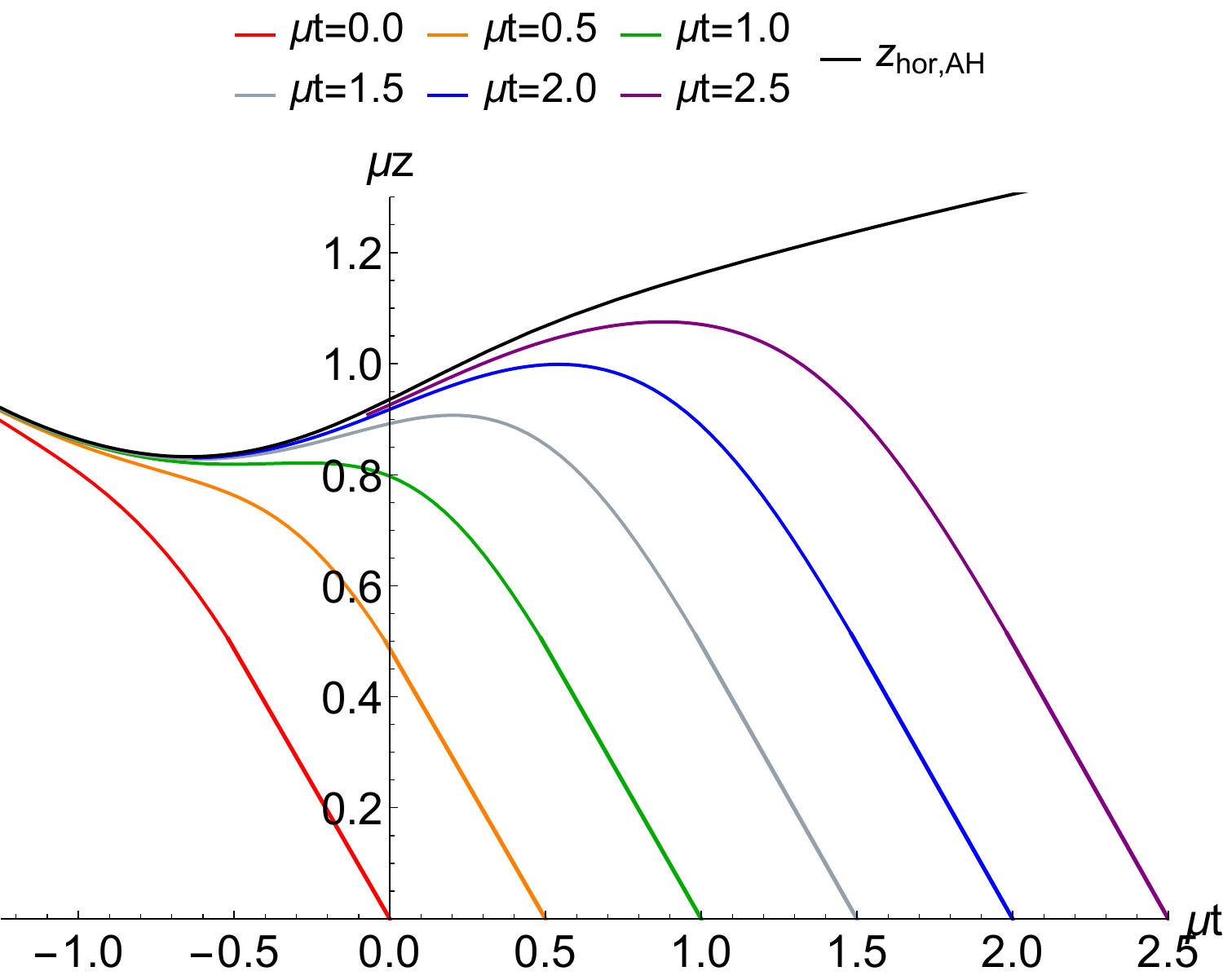}$\;$\includegraphics[width=5.5cm]{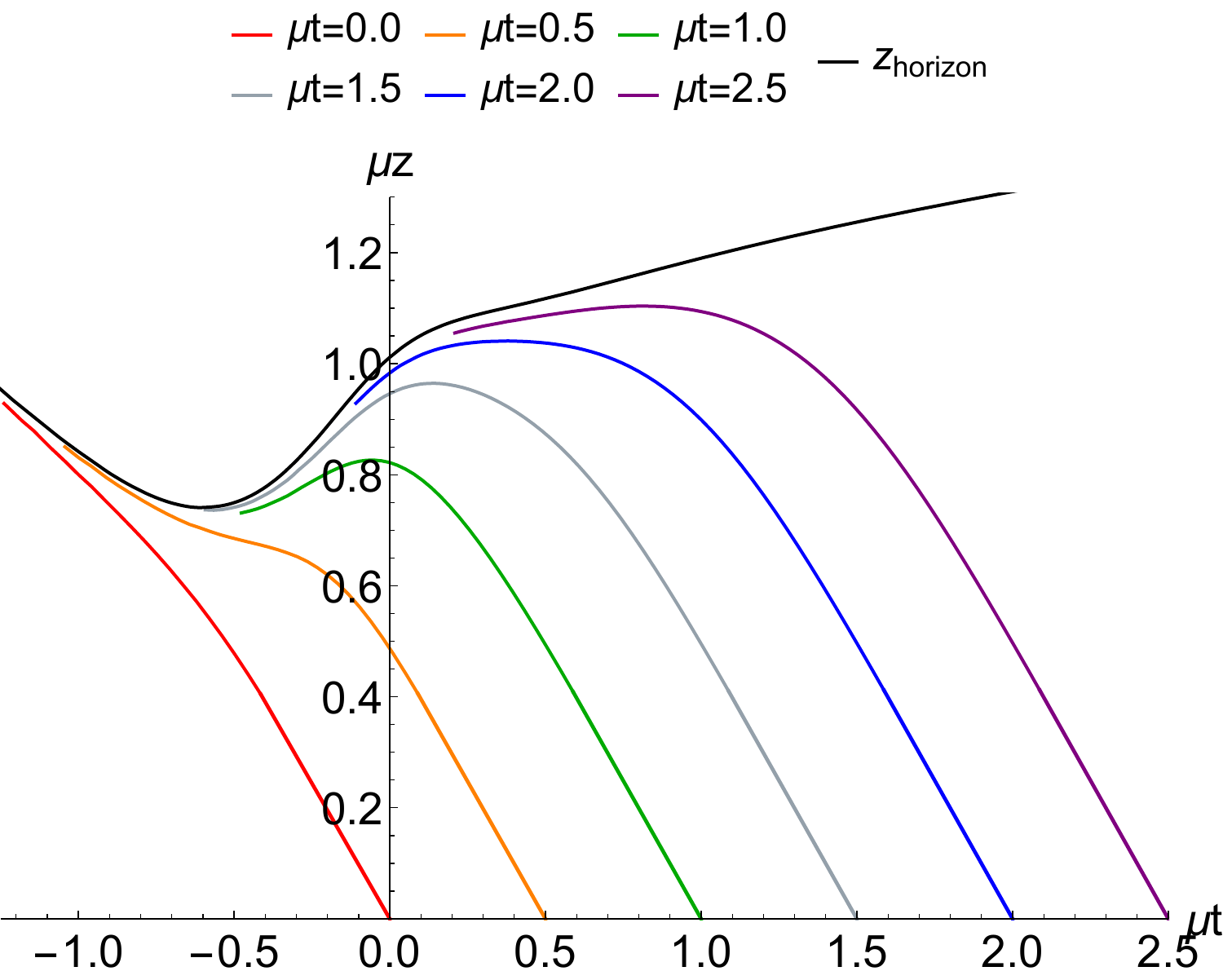}$\;$\includegraphics[width=5.5cm]{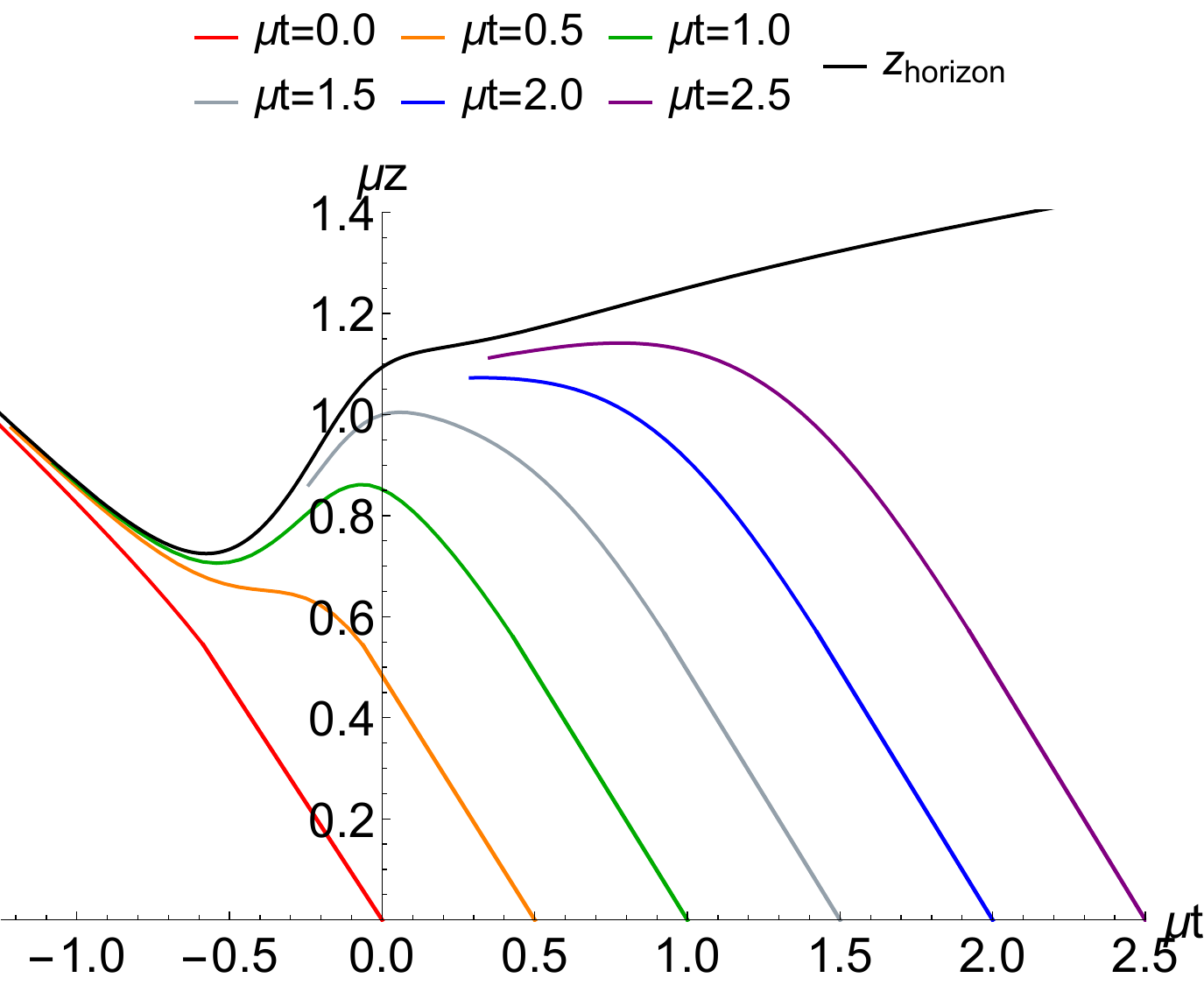}
\caption[Conformal geodesics used for holographic entanglement entropy.]{\label{GeodesicEEDip} 
The $z$-position of the geodesics at $y=0$ for several times and separations, starting with $l=0$ near the boundary, and increasing going towards the end of the curve. We show wide, intermediate and narrow shocks (from left to right). The $z$-position of the apparent horizon at $y=0$ is shown in black. In all the cases we studied the geodesics do not cross the horizon.}
%\end{center}
\end{figure}

\begin{table}[t]  \label{fit}
 \caption[Late time fit of entanglement entropy.]{Late time fit of the EE in the time range $\mu t\in[2.0,6.0]$.}
 \centering
 \begin{tabular}{lcccccc} 
 \hline\hline
  $\mu l$ & $a_w$ & $a_i$ & $a_n$ & $b_w$ & $b_i$ & $b_n$ \\
  \hline
  0.5     & 0.202 & 0.171 & 0.158 & -1.136 & -0.978 & -1.074 \\
  1.0     & 0.709 & 0.602 & 0.564 & -1.092 & -0.961 & -1.035 \\
  1.5     & 1.276 & 1.099 & 1.031 & -1.036 & -0.952 & -0.982 \\
 \hline
\end{tabular}
\end{table}

\begin{table}[t]  \label{fit1}
 \caption[Late time fit of the effective entropy density.]{Late time fit of the effective entropy density in the time range $\mu t\in[2.0,6.0]$.}
 \centering
 \begin{tabular}{lcccccc} 
 \hline\hline
  $\mu l$ & $a_w$ & $a_i$ & $a_n$ & $b_w$ & $b_i$ & $b_n$ \\
  \hline
  0.5     & 1.042 & 0.696 & 0.665 & -1.107 & -0.749 & -0.952 \\
  1.0     & 2.035 & 1.430 & 1.372 & -1.088 & -0.766 & -0.971 \\
  1.5     & 2.924 & 2.244 & 2.241 & -1.054 & -0.795 & -1.027 \\
 \hline
\end{tabular}
\end{table}

Let us now discuss the results from the evolution in the separation. The geometrical setup and the evolution in the separation at different times are shown in Fig.~\ref{EELevo}.
Analogous to Fig. \ref{GeodesicDip} we show in Fig. \ref{GeodesicEEDip} again the position of the tip of the extremal surface, this time for the EE.
Surprisingly, contrary to case of the 2-point function we never see the tip crossing the horizon, and in fact it always closely follows the horizon for larger separations. This is again perhaps counter-intuitive, since one would usually think about the EE as a more `nonlocal' quantity than the 2-point functions, and hence probing deeper in the bulk. Indeed, this is the case for pure AdS and also for thermal AdS, but in this case for large enough separations the 2-point function at the same time and length probes deeper in the bulk than the EE.

Of course our simulations only probed a limited set of times and lengths for our extremal surfaces and hence we cannot make a general statement if the EE never probes beyond the apparent horizon in geometries produced by shock wave collisions. Nevertheless, we think we have strong evidence that this is so, mainly since increasing the lengths at our chosen times clearly moves the tip of the surface along the horizon. We furthermore checked that extremal surfaces centered around $y \neq 0$ behave similarly, so that the property is not due to our symmetric set-up.

\section{Conclusions}\label{se:5}

In the paper at hand we studied the time evolution of equal time 2-point functions and HEE in strongly coupled anisotropic and inhomogeneous $\mathcal{N}=4$ super Yang Mills theory via its dual description.
In the dual description this amounts to finding geodesics and extremal surfaces in the gravitational background of two colliding gravitational  shock waves.
We used three different initial conditions, corresponding to wide, intermediate and narrow shocks.  

When the separation is held fixed the 2-point functions decrease before and increase after  the collision.
During the collision new correlations form such that the system becomes more correlated than in the beginning.
The narrower the shocks the higher the gain in correlations before they reach their final value. 

We also studied the correlation between the two shocks itself by following the maximum of the energy density. In this case the correlation between the two shocks increases linearly before the collision.
After the collision correlations  decrease for wide shocks, whereas for the narrower shocks they continue to grow before they fall off again. 

The time evolution of the EE can be divided into four regimes, namely highly efficient rapid initial growth, linear growth, post collisional regime and late time fall off. 
The smaller the shocks the more rapid the initial growth, reflecting the fact that the rate at which the energy density enters the entangling region is larger for smaller shocks. 
The post collisional regime is qualitatively different for the different initial conditions. As the shocks get smaller an additional minimum appears which we attribute to the fact that the longitudinal pressure becomes negative. The existence or absence of a minimum in EE in the post collisional regime thus serves as an order parameter to discriminate between the transparency (narrow shocks) and full-stopping (wide shocks) scenarios. 
At late times we observe polynomial fall off behavior where the exponent depends on the initial conditions. 

Surprisingly, we found that 2-point functions can probe behind the horizon, but only after the system has hydrodynamized. In contrast, the EE surface did not probe behind the horizon in our simulations, which is perhaps counter-intuitive.

 This finding has to be contrasted to the observations made in \cite{AbajoArrastia:2010yt}, where the authors studied the holographic entanglement entropy in Vaidya $AdS_3$ and found geodesics which cross the apparent horizon. In $AdS_3/CFT_2$, however,  the holographic entanglement entropy and the two-point function are equivalent, whereas in our $AdS_5$ they have manifestly different behavior. 

An interesting application of our results is to check numerically the quantum null energy condition \cite{Bousso:2015mna,Bousso:2015wca,Koeller:2015qmn} in a regime where the classical null energy condition breaks down. Namely, for the narrow shock-waves shortly after the collision there are regions where the classical null energy condition fails. We intend to perform this check in future work using the results for HEE established in the present work.

An interesting generalization of our results could be the consideration of shock-wave collisions in non-conformal theories, holographically modeled by the addition of a scalar field with judiciously chosen self-interactions \cite{Attems:2016ugt,Attems:2016tby}.

\section*{Acknowledgments}

We thank Maximilian Attems, Hajar Ebrahim, Ognen Kapetanoski, Behnoush Khavari, Hong Liu, Esperanza Lopez, Eugenio Megias, Ayan Mukhopadhyay, Florian Preis, Toni Rebhan, Paul Romatschke and Andrei Starinets for discussions. 

DG thanks the Center for Theoretical Physics at the Massachusetts Institute of Technology for hospitality while part of this work was completed. 

This work was supported by the following projects of the Austrian Science Fund (FWF): Y435-N16, I952-N16, P27182-N27, P28751-N27, DKW1252-N27 and P26328. WS is supported
by the U.S. Department of Energy under grant Contract Number DE-SC0011090.

%%%%%%%%%%%%%%%%%%%%%%%%%%%%%%%%%%%%%%%%%%%%%
\begin{appendix}
%%%%%%%%%%%%%%%%%%%%%%%%%%%%%%%%%%%%%%%%%%%%%
\section{Near boundary expansion of the shock wave spacetime}\label{app:1}
Here we work in a gauge where we exploit the residual gauge freedom to set $\xi(v,y)=0$. 
In this gauge the near boundary expansion of the shockwave metric up to $6^{\rm th}$ order in z is given by

\bse\label{asympMetric}
\ba
A(z,t,y)&=&\frac{1}{z^2}+z^2 a_4+\frac{1}{2} z^3\partial_t a_4\nonumber\\ 
        &+&\frac{1}{20}z^4\Big(3\partial_t^2 a_4-\partial_y^2 a_4+4 \partial_y^2 b_4\Big)+\mathcal{O}(z^{7})\\
B(z,t,y)&=&z^4 b_4+z^5\Big(\partial_t b_4 +\frac{2}{15}\partial_y f_4\Big)\nonumber\\
        &+&\frac{1}{180}z^6\Big(4\partial_y^2 a_4 +5\partial_y^2 b_4+105\partial_t^2 b_4+30 \partial_t \partial_y f_4\Big)+\mathcal{O}(z^7)\\
S(z,t,y)&=&\frac{1}{z}+z^4\Big(-\frac{1}{20}\partial_t a_4-\frac{1}{15} \partial_y f_4 \Big)\nonumber\\
        &+&\frac{1}{180}z^5\Big(\partial_y^2 a_4-3\partial_t^2 a_4 +8 \partial_y^2 b_4\Big)+\mathcal{O}(z^{7})\\ 
F(z,t,y)&=&z^2 f_4 +\frac{1}{5}z^3\Big(\partial_y a_4 +4 \partial_t f_4\Big)\nonumber\\
        &+&\frac{1}{6}z^4\Big(\partial_t\partial_y a_4-\partial_t\partial_y b_4+ 2\partial_t^2 f_4\Big)+\mathcal{O}(z^{7})\,.
\ea
\ese

\section{Numerical checks}\label{app:2}

\begin{figure}
\begin{center}
\includegraphics[scale=0.35]{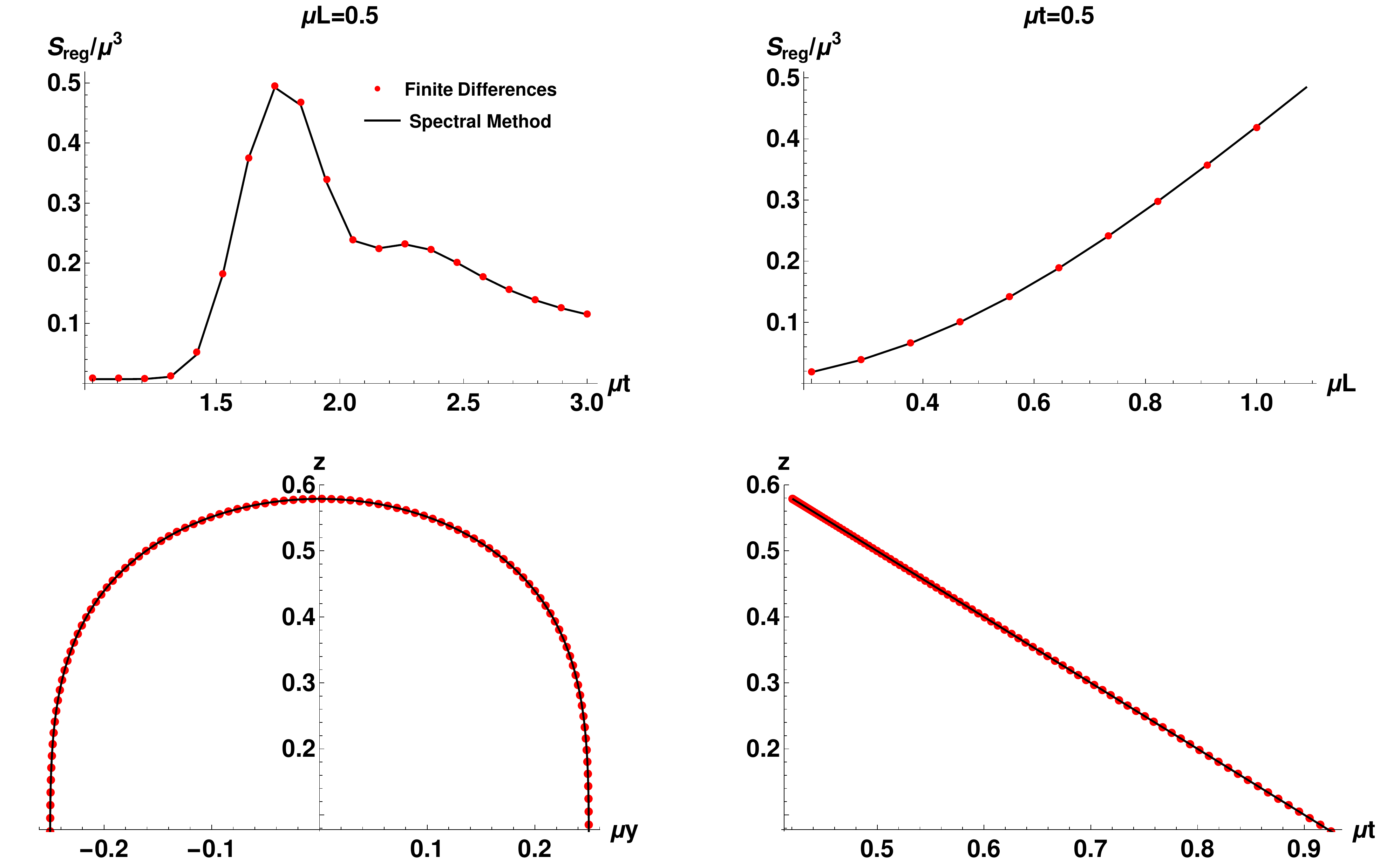}
\caption[Comparison of both numerical codes.]{\label{CheckCode} 
Comparison of results from the relaxation algorithm using a first order finite difference method (red dots) and a spectral method (black lines). On top we show the time evolution (left) and the scaling with the separation (right) of the regularized EE for narrow shocks. 
On the bottom we show a conformal geodesic in the y-z plane (left) and the t-z plane (right).
 }
\end{center}
\end{figure}

In any numerical analysis it is important to check the underlying algorithm for programming mistakes and to track numerical errors.
In order to check the correctness of our numerical results two completely independent relaxation codes were developed, one by the Vienna group and another one by Wilke van der Schee. The first algorithm employs first order finite differences, the second one a spectral method. We find excellent agreement (see Fig.~\ref{CheckCode}).

In both computer codes the embedding functions of the geodesics are represented on a finite number of grid points. The numerical result must converge to the true solution when the number of gridpoints is increased. Table \ref{gridsizeTable} demonstrates that both, the 2-point function and the EE, change only insignificantly already for a moderate number of 200 grid points. Based on this analysis we have chosen 200 gridpoints in all our simulations.

\begin{table}[h]  \label{gridsizeTable}
 \caption[Scaling of non-local observables with number of gridpoints.]{Scaling of the 2-point function $e^{-L_{\textrm{\tiny reg}}}$ and the EE $S_{\textrm{\tiny reg}}$ with the number of gridpoints.
 The results are for narrow shocks at the collision time ($\mu t=0$) and at some later time ($\mu t=2$). For the 2-point function the separation is $\mu l=1.0$ and for the EE $\mu l=0.5$. In both cases the cutoff is fixed at $z_{\textrm{\tiny cut}}=0.075$.}
 \centering\bigskip
 \begin{tabular}{lccccc} 
 \hline\hline
   gridsize & $e^{-L_{\textrm{\tiny reg}}}|_{\mu t=0}$ & $e^{-L_{\textrm{\tiny reg}}}|_{\mu t=2}$ & $S_{\textrm{\tiny reg}}|_{\mu t=0}$ & $S_{\textrm{\tiny reg}}|_{\mu t=2}$\\
  \hline
   50       & 0.907817      & 0.997721     & 0.468440       & 0.0369731\\
   80       & 0.908684      & 0.998197     & 0.496564       & 0.0712919\\
   100      & 0.908881      & 0.998306     & 0.498700       & 0.0735928\\
   200      & 0.909140      & 0.998450     & 0.500354       & 0.0744153\\
   300      & 0.909187      & 0.998476     & 0.500660       & 0.0744176\\
   400      & 0.909204      & 0.998486     & 0.500772       & 0.0744166\\
 \hline
\end{tabular}
\end{table}

\begin{table}[h]  \label{zcutTable}
 \caption[Scaling of non-local observables with cutoff.]{Scaling of the 2-point function $e^{-L_{\textrm{\tiny reg}}}$ and the EE $S_{\textrm{\tiny reg}}$ with the cutoff $z_{\textrm{\tiny cut}}$.
 The results are for narrow shocks at the collision time ($\mu t=0$) and at some later time ($\mu t=2$). For the 2-point function the separation is fixed to $\mu l=1.0$ and for the EE to $\mu l=0.5$. In both cases 200 gridpoints are used.}
 \centering
 \begin{tabular}{lccccc} 
 \hline\hline
  $z_{\textrm{\tiny cut}}$ & $e^{-L_{\textrm{\tiny reg}}}|_{\mu t=0}$ & $e^{-L_{\textrm{\tiny reg}}}|_{\mu t=2}$ & $S_{\textrm{\tiny reg}}|_{\mu t=0}$ & $S_{\textrm{\tiny reg}}|_{\mu t=2}$\\
  \hline
   0.1       & 0.909028     & 0.998464   & 0.504097       & 0.0747103\\
   0.09      & 0.909079     & 0.998458   & 0.502534       & 0.0746000\\
   0.08      & 0.909122     & 0.998453   & 0.501073       & 0.0744817\\
   0.07      & 0.909156     & 0.998448   & 0.499622       & 0.0743396\\
   0.06      & 0.909181     & 0.998444   & 0.498010       & 0.0741270\\
   0.05      & 0.909195     & 0.998440   & 0.495843       & 0.0736603\\
   0.04      & 0.909191     & 0.998436   & 0.491721       &          \\
   0.03      & 0.909157     & 0.998432   &                &          \\
   0.02      & 0.909035     & 0.998428   &                &          \\
   0.01      & 0.908378     & 0.998470   &                &          \\
 \hline
\end{tabular}
\end{table}

Our numerical scheme employs a cutoff $z_{\textrm{\tiny cut}}$ in the holographic coordinate.
The final result for our observables should not depend on this cutoff which purely serves numerical purposes. 
In Table \ref{zcutTable} we show the scaling of the 2-point function of separation $\mu l=1$ and the EE of separation $\mu l=0.5$ evaluated at two different times ($\mu t=0,2$) for the narrow shocks. 
The results for the 2-point function are nicely independent of the cutoff in the range $z_{\textrm{\tiny cut}}\in[0.01,0.1]$.
In case of the EE the cutoff dependence turns to be $\approx 1 \%$ in the range  $z_{\textrm{\tiny cut}}\in[0.05,0.1]$ which is sufficient for our qualitative studies. In all our simulations presented in this work we have set the cutoff to $z_{\textrm{\tiny cut}}=0.075$.

%\section{Tracking a single shock}\label{app:4}
%
%In addition to the previously obtained results we can also study how a single shock gets entangled with its surroundings  throughout the collision by tracking the maximum of the energy density of a single shock which for intermediate and narrow shocks corresponds to following the light cone.
%\begin{figure}
%\begin{minipage}[t]{0.5\textwidth}
%%\vspace{-0.8cm}
%%\hspace{-0.5cm}
%\includegraphics[width=0.9\textwidth]{EEFollowWide.jpg}
%\end{minipage}
%\begin{minipage}[t]{0.5\textwidth}
%%\vspace{0.1cm}
%\includegraphics[width=0.9\textwidth]{EEFollowNarrow.jpg}
%\end{minipage}
%\begin{minipage}[t]{0.5\textwidth}
%%\vspace{0.1cm}
%%\hspace{-0.5cm}
%\includegraphics[width=0.9\textwidth]{EEFollowVeryNarrow.jpg}
%\end{minipage}
%\begin{minipage}[t]{0.5\textwidth}
%%\vspace{0.1cm}
%\includegraphics[width=\textwidth]{GeoEEFollowNarrow.jpg}
%\end{minipage}
%\caption{Evolution of EE by tracking a single shock for wide, intermediate and narrow shocks (from left to right).
% }\label{EEfollow}
%\end{figure}
%The shape  of the curves look quite similar to the results obtained in the  previous section.
%The difference is  in the fall off behavior, which is significantly faster than for the mid rapidity region and drops below the initial entanglement which never happens in the mid rapidity case. 
%This can be understood from a particle perspective. As more and more particles escape from the lightcone and spread out, less particles can be entangled. This is also reflected in the energy density which after the collision is below the initial value. 

\end{appendix}

\bibliographystyle{JHEP-2} 

\bibliography{review,shocks}

%\providecommand{\href}[2]{#2}\begingroup\raggedright\begin{thebibliography}{10}
%\addcontentsline{toc}{section}{References}

%\end{thebibliography}\endgroup

\end{document}